\documentclass[11pt]{article}
\usepackage{amssymb}
\usepackage{amsmath}
\usepackage[all]{xy}
\usepackage[dvips]{graphicx}
\numberwithin{equation}{section}

\def\be{\begin{equation}}
\def\ee{\end{equation}}
\def\ba{\begin{align}}
\def\ea{\end{align}}
\def\beq{\begin{eqnarray}}
\def\eeq{\end{eqnarray}}

\def\p{\partial}

\def\zbar{\bar z}

 \input{epsf}

 \usepackage{epsfig}

\begin{document}

\title{\Large{\bf 
The conformal current algebra on supergroups with applications
to the spectrum and integrability
}} 
\author{Raphael Benichou$^{a}$ and Jan Troost$^{b}$ 
}
\date{}
\maketitle
\begin{center}
  $^{a}$ Theoretische Natuurkunde, Vrije Universiteit Brussel,\\
 Pleinlaan 2, B-1050 Brussels, Belgium \\
  \vspace{.3cm}
  $^{b}$Laboratoire de Physique Th\'eorique \\
Unit\'e Mixte du CRNS et
    de l'\'Ecole Normale Sup\'erieure \\ associ\'ee \`a l'Universit\'e Pierre et
    Marie Curie 6 \\ UMR
   8549  \footnote{Preprint LPTENS-10/11.} 
\\ \'Ecole Normale Sup\'erieure \\
  $24$ Rue Lhomond Paris $75005$, France
\end{center}

 \begin{abstract}
   We compute the algebra of left and right currents for a principal
   chiral model with arbitrary Wess-Zumino term on supergroups with
   zero Killing form.  We define primary fields for the current
   algebra that match the affine primaries at the Wess-Zumino-Witten
   points.  The Maurer-Cartan equation together with current
   conservation tightly constrain the current-current and
   current-primary operator product expansions.
   The Hilbert space of the theory is generated by acting with the
   currents on primary fields.  We compute the conformal dimensions of
   a subset of these states in the large radius limit. 
   The current
   algebra is shown to be consistent with the quantum integrability of
   these models to several orders in perturbation theory.
\end{abstract}

\newpage

\tableofcontents

\newpage

\section{Introduction}
Principal chiral models with Wess-Zumino term on supergroups (and
their cosets) arise in many contexts including string theory on
Anti-de Sitter backgrounds with Ramond-Ramond fluxes, the integer
quantum hall effect, quenched disorder systems, polymers, as well as
other domains in physics. When the supergroup
has zero Killing form, the model is perturbatively conformal
\cite{Berkovits:1999im, Bershadsky:1999hk, Babichenko:2006uc}. Thus,
these models provide us with a two-parameter family of two-dimensional
conformal field theories with supergroup symmetry. They exhibit a
current algebra which is conformal and non-chiral \cite{Ashok:2009xx}.
Since these models fall into a class which exhibits integrability at
least classically and most likely quantum mechanically, these
two-dimensional conformal field theories may allow for an exact
determination of their spectrum.

Steps towards solving these models were made using various techniques.
For particular supergroups the Wess-Zumino-Witten points are
well-understood
\cite{Rozansky:1992rx}\cite{Schomerus:2005bf}\cite{Gotz:2006qp}.  The
bulk spectrum was computed at some special points of the moduli space
in \cite{Read:2001pz}.  The spectrum for states living on particular
boundaries can be obtained at any point of the moduli space
\cite{Quella:2007sg, Mitev:2008yt,Candu:2009ep}.  Methods to compute
a subset of correlation functions were recently proposed in
\cite{Candu:2010yg}.
Despite these successes, the determination of the full bulk 
spectrum of the 
conformal field theories on supergroups remains an open problem.

A strong motivation for determining the spectrum of these models, and
their cosets, is the prospect of solving string theory in $AdS$
space-times in conformal gauge, which via holography \cite{Maldacena:1997re}
may lead to a
neater formulation of the solution of gauge theories at large $N$ \cite{Gromov:2009tv}.
Our attitude in attacking this problem is to first attempt to solve
for the spectrum in conformal gauge on supergroups (relevant to $AdS_3$ string theory for
instance), and then for the spectrum on supercosets (relevant to
$AdS_5$ string theory for example).

In this paper, we take a further step in our understanding of the symmetry,
the integrability and the Hilbert space and spectrum of these models.
In section \ref{worldsheetcurrentalgebra}, we review the conformal
current algebra \cite{Ashok:2009xx} obeyed by the conserved current
associated to the left action of the supergroup on itself. We will
determine further terms at order zero in the current algebra.  In
section \ref{left-right} we compute the interplay between the left and
the right conformal current algebra, as well as with the adjoint
primary operator.  In section \ref{primaries} we define the primary
fields for the current algebra.  These fields correspond to the affine
primaries at the Wess-Zumino-Witten points. 
We show that current primaries are also Virasoro primaries and compute
their conformal dimension at large radius.  
In section \ref{bootstrap} we explain how to compute the 
current-current and current-primary OPEs
  order by order 
in perturbation theory by demanding consistency with 
current conservation and the Maurer-Cartan equation.
In section
\ref{confdimcomp} we compute conformal dimensions of operators that
are composites of a current and a primary to first order in
semi-classical perturbation theory.  We argue that the Hilbert space
is generated by composites of currents and primary fields and show
how to compute the conformal dimension of such operators 
in semi-classical perturbation theory.
In section \ref{integrability}
we comment on the classical and quantum integrability of the model,
and its consistency with the conformal current algebra.  We conclude
in section \ref{conclusions}.

We have gathered many technical details in the appendices.  In appendix
\ref{compositeOPEs} we give a prescription to compute OPEs involving
composite operators.  In appendix \ref{XXOPEs} we compute the behavior
at large radius for the coefficients appearing in the current-current
and current-primary OPEs.  In appendix \ref{consistentPertOPEs} we 
prove the consistency of the 
perturbative 
algorithm used to compute the 
current-current and current-primary OPEs.
Appendix \ref{AppCurrents} contains further
consistency checks of the current algebra as well as details of the
computation of the current algebra.  In appendix \ref{AppPrimaries} we
detail calculations involving the primary fields.  In
appendix \ref{commutators} we translate the current-current OPEs into
(anti-)commutation relations for the modes of the currents when the
theory is defined on a cylinder.  Finally, classical integrability of
the model is proven in appendix \ref{classint}

\section{The conformal current algebra}
\label{worldsheetcurrentalgebra}
\subsection*{Setting}
We study a non-linear sigma-model on a supergroup $G$ with zero
Killing form, including a kinetic term and a Wess-Zumino term with
arbitrary coefficient. The model is conformal and has a global
symmetry group corresponding to the left and right action of the group
on itself. 
In this section we review and complement the analysis of the algebra
of current components associated to the left group action
\cite{Ashok:2009xx}. 
 The action of the non-linear sigma-model on the
supergroup is:
\begin{align}\label{ourmodel}
S &= S_{kin} + S_{WZ}\cr
S_{kin} &=  \frac{1}{ 16 \pi f^2}\int d^2 z Tr'[- \partial^\mu g^{-1}
\partial_\mu g]
\cr
S_{WZ} &= - \frac{ik}{24 \pi} \int_B d^3 y \epsilon^{\alpha \beta \gamma}
Tr' (g^{-1} \partial_\alpha g g^{-1} \partial_\beta g   g^{-1} \partial_\gamma g )
\end{align}
where $g$ takes values in the supergroup $G$ and $Tr'$ indicates the
non-degenerate bi-invariant metric. We will use the normalization and
results of \cite{Ashok:2009xx}. The Wess-Zumino-Witten points are
given by the equation $1/f^2 = |k|$.  Note that the action is
invariant under group inversion $g \leftrightarrow g^{-1}$ and
simultaneous orientation reversal $z \leftrightarrow \bar{z}$.

\subsection*{The conformal current algebra}
{From} the action \eqref{ourmodel} we can calculate the classical
currents associated to the invariance of the theory under left
multiplication of the field $g$ by a group element in $G_L$ and right
multiplication by a group element in $G_R$.
The classical $G_L$ currents are given by
\begin{align}\label{normeqn}
j_{L,z} &= c_+ \partial g g^{-1}\cr
j_{L,\bar{z}} &= c_- \bar{\partial} g g^{-1} \,,
\end{align}
where the constant $c_+$ and $c_-$ are given in terms of the couplings
by: \be\label{c+-} c_{\pm} = -\frac{(1\pm kf^2)}{2f^2} \,.  \ee
Similarly, we also have the left-invariant currents that generate
right multiplication:
\begin{align}
j_{R,z} &= -c_- g^{-1} \p g \cr
j_{R,\bar z} &= -c_+ g^{-1} \bar \p g\, .
\end{align}
The operator product expansions (OPEs) satisfied by the left currents have been
derived in \cite{Ashok:2009xx}. They read:
\begin{align}\label{euclidOPEs}
j_{L,z}^a (z) &j_{L,z}^b (0)  \sim  \ \kappa^{ab} \frac{c_1}{z^2} 
  + {f^{ab}}_c \left[ \frac{c_2}{z} j_{L,z}^c(0)+ (c_2-g) \frac{\bar{z}}{z^2} j_{L,\bar{z}}^c(0) \right] \cr
& +  {f^{ab}}_c \left[-\frac{g}{4}\frac{\bar z}{z}(\partial_z j_{\bar z}^c(0)-\partial_{\bar z}j_z^c(0)) +  \frac{c_2}{2} \partial_z j_{L,z}^c(0)+ \frac{c_2-g}{2} \frac{\bar{z}^2}{z^2} \partial_{\bar z}j_{L,\bar{z}}^c(0) \right] \cr
& + :j_z^a j_z^b:(0) 
+ {{A}^{ab}}_{cd}  \frac{1}{2} \frac{\bar z^2}{z^2}:j_{\bar z}^{ c} j_{\bar z}^{d }:(0) + {{B}^{ab}}_{cd}  \frac{\bar z}{z} :j_z^{ c} j_{\bar z}^{d }:(0) -
 {{C}^{ab}}_{cd}  \log |z|^2 :j^{ c}_z j_z^{e } :(0)  \cr
& + ... \cr
j_{L,\bar{z}}^a (z)& j_{L,\bar{z}}^b (0) \sim  \ \kappa^{ab} c_3 \frac{1}{\bar{z}^2} 
  + {f^{ab}}_c \left[  \frac{c_4}{\bar{z}} j_{L,\bar{z}}^c(0) + \frac{(c_4-g)z}{\bar{z}^2} j_{L,z}^c(0)\right] \cr
& + {f^{ab}}_c \left[\frac{g}{4}\frac{z}{\bar z}(\partial_z j_{\bar z}^c(0)-\partial_{\bar z}j_z^c(0)) +  \frac{c_4}{2} \partial_{\bar z} j_{L,\bar z}^c(0)+ \frac{c_4-g}{2} \frac{z^2}{\bar z^2} \partial_{z}j_{L,z}^c(0) \right] \cr
& + :j_{\bar z}^a j_{\bar z}^b:(0)  
- {{A}^{ab}}_{cd} \log |z|^2 :j_{\bar z}^{ c} j_{\bar z}^{d }:(0) +  {{B}^{ab}}_{cd}\frac{ z}{\bar z} :j_z^{ c} j_{\bar z}^{d }:(0) + {{C}^{ab}}_{cd} \frac{1}{2} \frac{z^2}{\bar z^2} :j^{ c}_z j_z^{d }:(0)  \cr
& + ... \cr
j_{L,z}^a (z) &j_{L,\bar{z}}^b(0) \sim \ \tilde{c}\kappa^{ab} 2\pi \delta^{(2)}(z-w) 
  + {f^{ab}}_c  \left[  \frac{(c_4-g)}{\bar{z}} j_{L,z}^c(0) + \frac{(c_2-g) }{z} j_{L,\bar{z}}^c(0) \right] \cr
& + {f^{ab}}_c  \left[ -\frac{g}{4} \log |z|^2 (\partial_z j_{\bar z}^c(0)-\partial_{\bar z}j_z^c(0)) + \frac{(c_4-g)z}{\bar{z}}\partial_z j_{L,z}^c(0) \right] \cr
& + :j_z^a j_{\bar z}^b:(0) 
+ {{A}^{ab}}_{cd} \frac{\bar z}{z}:j_{\bar z}^{ c} j_{\bar z}^{d }:(0) - {{B}^{ab}}_{cd} \log |z|^2 :j_z^{ c} j_{\bar z}^{d }:(0) 
+ {{C}^{ab}}_{cd} \frac{z}{\bar z} :j^{ c}_z j_z^{d }:(0)  \cr
& + ... \cr
\end{align}
Compared to \cite{Ashok:2009xx}, we have added a few terms at order
zero in the distance between the insertion points of the two current
components\footnote{We would like to thank Anatoly Konechny for
  stressing the importance of these terms, and for sharing his
  insights in these terms in perturbation theory near
  Wess-Zumino-Witten points.}.  The
ellipses
refer to subleading terms in the expansion in the distance between the
two insertion points (which includes logarithms).  The right current
components $j_{R,z}$ and $j_{R,\zbar}$ satisfy similar operator
product expansions amongst themselves, with the holomorphic
coordinates replaced by anti-holomorphic ones. This can be proven by
using the $\mathbb{Z}_2$ symmetry that we noted before.
Associativity of the current algebra is discussed in appendix \ref{associativity}.

For the supergroup non-linear
sigma-model in equation \eqref{ourmodel}, the coefficients of the
second and first order poles in the
conformal current algebra, expressed purely in terms of $c_{\pm}$, are
given by \cite{Ashok:2009xx}
\begin{align}\label{candg}
c_1 &= -\frac{c_+^2}{c_++c_-} \qquad\qquad\qquad c_3 = -\frac{c_-^2}{c_++c_-} \cr
c_2 &= i \frac{c_+(c_++2c_-)}{(c_++c_-)^2} \qquad\qquad c_4 =  i \frac{c_-(2c_++c_-)}{(c_++c_-)^2}  \cr 
g &= i \frac{2c_+c_-}{(c_++c_-)^2} \qquad\qquad\qquad \tilde{c} = \frac{ c_+ c_-}{c_++c_-} \,,
\end{align}
where the coefficients $c_{\pm}$ are the factors defined in 
equation \eqref{c+-} as the normalization of the currents.
The coefficients $c_i$ are exact.

The current algebra defined by equation \eqref{euclidOPEs} is
compatible with both current conservation and the Maurer-Cartan
equation :
\be\label{CC} \bar \partial j_{L,z}^a + \partial j_{L,\bar z}^a = 0 \ee
\be\label{MC} c_- \bar \partial j_{L,z}^a - c_+ \partial j_{L,\bar z}^a - i {f^a}_{bc} :j_{L, z}^c  j_{L,\bar z}^b : = 0. \ee
Indeed the OPE of a current with the left-hand side of the current
conservation equation \eqref{CC} (respectively the Maurer-Cartan
equation \eqref{MC}) gives zero up to contact terms (respectively
exactly zero).  Moreover, demanding compatibility of the current
algebra with both equations \eqref{CC} and \eqref{MC} is a way to
determine all the other subleading terms in the current algebra, order
by order in semi-classical perturbation theory, namely for small
$f^2$ (at fixed $kf^2$).  This is explained in section \ref{bootstrap}.
As we will see, the assumption of the validity of current conservation and
especially the Maurer-Cartan equation in the quantum theory,
determines a tightly constrained
and interesting algebraic structure associated to supergroups with
vanishing Killing form.  This hypothesis is tightly linked to the
quantum integrability of the model, as we discuss in section
\ref{integrability}.

We can use this perturbative technique to compute
the coefficients of the current bilinears that appear in
equation \eqref{euclidOPEs}, up to order
$f^2$. This computation is detailed in appendix \ref{jMCOPE} and it leads to
the results:
\beq {A^{ab}}_{cd} &=& \frac{c_+^2}{(c_++c_-)^3} \frac{1}{2}
 ({f^b}_{cg} {f^{ag}}_d (-1)^{cd} +  {f^b}_{dg} {f^{ag}}_c)
 +\mathcal{O}(f^4)\cr
 {B^{ab}}_{cd} &=&
 \frac{c_+c_-}{(c_++c_-)^3} ({f^b}_{cg} {f^{ag}}_d (-1)^{cd} +  {f^b}_{dg} {f^{ag}}_c)
+ \mathcal{O}(f^4)\cr
 {C^{ab}}_{cd} &=&
\frac{c_-^2}{(c_++c_-)^3}\frac{1}{2} ({f^b}_{cg} {f^{ag}}_d (-1)^{cd} +  {f^b}_{dg} {f^{ag}}_c)+\mathcal{O}(f^4). 
\label{ABC}
\eeq 
The fact that the same tensors appear in the three different
current-current OPEs \eqref{euclidOPEs} is a consequence of current
conservation.  The four-tensors $A,B,C$ are (graded) symmetric in
their two upper indices. This follows from the interchangeability of
the current components on the left hand side of the first two OPEs in
\eqref{euclidOPEs}.  Equation \eqref{ABC} shows that these
four-tensors are also (graded) symmetric in their two lower indices.
Thus they are linear maps from graded symmetric tensors onto
graded symmetric tensors. 
 They partially code higher order corrections
to equation \eqref{ABC} (see appendix \ref{jMCOPE}).

In appendix \ref{jMCOPE}, we have included a careful discussion of
minus signs arising due to the graded statistics of the
supergroup. For the remainder of the paper however, we will not be
careful about minus signs arising due to the grading of
operators. Since we use only universal group and (super) Lie algebra
properties in our calculations, all signs can be consistently restored.

\subsection*{The Virasoro algebra from the current algebra}
In \cite{Ashok:2009xx} it was shown that the left and right Virasoro algebra emerge from the current algebra \eqref{euclidOPEs} via the Sugawara construction. For instance the holomorphic stress-tensor :
\be T(z) = \frac{1}{2 c_1} \kappa_{ba} :j^a_{L,z} j^b_{L,z}: \ee
satisfies the following OPEs :
\be\label{Tj} T(z) j^a_{L,z}(w) = \frac{j^a_{L,z}(w)}{(z-w)^2} +  \frac{\p j^a_{L,z}(w)}{z-w} + \mathcal{O}(z-w)^0 \ee
\be\label{Tjbar} T(z) j^a_{L,\bar z}(w) =   \frac{\p j^a_{L,\bar z}(w)}{ z- w} + \mathcal{O}(z-w)^0 \ee
\be\label{TT} T(z) T(w) = \frac{sdim(G)}{2(z-w)^4} + \frac{T(w)}{(z-w)^2} +  \frac{\p T(w)}{z-w} + \mathcal{O}(z-w)^0. \ee
In appendix \ref{TjandTT} we give more details of the proof of
equation \eqref{Tj}.  In particular it is shown that the terms of
order zero in equation \eqref{euclidOPEs} (as well as any of the other
subleading terms) do not modify this OPE.  
We also checked through explicit computation that the 
invariant contractions of the
structure constants and the metric with the four-tensors \eqref{ABC}
appearing in the energy-momentum tensor/current OPE give zero for
the algebra $psl(2|2)$.

\section{The left-right current algebra}
\label{left-right}
In this section, we compute the operator product expansions of currents 
associated to the left and the right action of the group upon itself.

\subsection{The primary adjoint operator}
The right current
components can be rewritten in terms of the adjoint group action on
the left currents:
\be\label{jLAdj=jR0}
j_{R,z} = -c_- g^{-1} \partial g = -\frac{c_-}{c_+} Ad_{g^{-1}} (j_{L,z})
\ee
\be\label{jLAdjB=jRB0} 
j_{R,\bar z} = -c_+ g^{-1} \bar \partial g = - \frac{c_+}{c_-} Ad_g (j_{L,\bar{z}}) .
\ee
In the quantum theory the adjoint group action is generated by an operator that we call the primary adjoint operator :
\be\label{PrimAdj} \mathcal{A}^{a \bar a}=  x \, Str (g^{-1} t^a g t^{\bar a}).\ee
Here $x$ is some normalization factor.
This operator transforms in the adjoint
representation with respect to both the left and the right
algebras. 
In the following unbarred (respectively barred) indices refer to the left (respectively
right) adjoint representation. We recall that this field is also 
useful in writing down the Lagrangian of the model, and that its anomalous
dimension is proportional to the beta-function of the model (which is
zero in the case under study) \cite{Knizhnik:1984nr}. 
Special properties of the primary adjoint operator 
in non-linear sigma models on supergroup with vanishing Killing form 
were also 
discussed in \cite{Quella:2007sg}. 
We can rewrite equations \eqref{jLAdj=jR0} and \eqref{jLAdjB=jRB0} as:
\be\label{jLAdj=jR} j_{R,z}^{\bar b} = -\frac{c_-}{c_+} \kappa_{ba}:j^a_{L,z} \mathcal{A}^{b \bar b}: \ee
\be\label{jLAdjB=jRB} j_{R,\bar z}^{\bar b} = -\frac{c_+}{c_-} \kappa_{ba}:j^a_{L,\bar z} \mathcal{A}^{b \bar b}:. \ee
Using the $\mathbb{Z}_2$ symmetry of the theory we have also:
\be\label{jRAdj=jL}  j^{b}_{L,z} = -\frac{c_+}{c_-} \kappa_{\bar b \bar a}:j^{\bar a}_{R,z} \mathcal{A}^{b \bar b}:(z)  \ee
\be\label{jRAdjB=jLB}  j^{b}_{L,\bar z} = -\frac{c_-}{c_+} \kappa_{\bar b \bar a}:j^{\bar a}_{R,\bar z} \mathcal{A}^{b \bar b}:(z) \ee
These relations fix 
a normalization for the operator $\mathcal{A}^{a \bar a}$. They are compatible if the following relations hold :
\be\label{AdjAdj=Ibar} \kappa_{ba} \mathcal{A}^{a \bar a} \mathcal{A}^{b \bar b} =
\kappa^{\bar a \bar b}I \ee \be\label{AdjAdj=I} \kappa_{\bar b \bar a}
\mathcal{A}^{a \bar a} \mathcal{A}^{b \bar b} = \kappa^{a b}I \ee
where $I$ is the identity at least as acting upon the current algebra.
One can argue more generically that these bilinears are proportional to the unit operator
by using the definition of the primary adjoint in terms of the
supertrace, and using completeness of the Lie algebra generators.
Remember also that the left and right conformal dimensions of the adjoint
operator $\mathcal{A}^{a \bar a}$ vanish since they are proportional to the
dual Coxeter number of the Lie superalgebra. 

The action of the zero modes of the currents generates the group transformations.
Since the structure constants are the generators of the Lie superalgebra in the adjoint representation, the OPE between a current and the primary adjoint operator reads :
\begin{align}\label{jAdj} & j^a_{L,z}(z) \mathcal{A}^{b \bar b}(w) = \frac{c_+}{c_++c_-} \frac{i{f^{ab}}_c \mathcal{A}^{c \bar b}}{z-w} + ... \cr
& j^a_{L,\bar z}(z) \mathcal{A}^{b \bar b}(w) 
= \frac{c_-}{c_++c_-} \frac{i{f^{ab}}_c \mathcal{A}^{c \bar b}}{\bar z- \bar w} +... \cr
& j^{\bar a}_{R,z}(z) \mathcal{A}^{b \bar b}(w) = \frac{c_-}{c_++c_-} \frac{i{f^{\bar a \bar b}}_{\bar c} \mathcal{A}^{b \bar c}}{z-w} +... \cr
& j^{\bar a}_{R,\bar z}(z) \mathcal{A}^{b \bar b}(w) = \frac{c_+}{c_++c_-} \frac{i{f^{\bar a \bar b}}_{\bar c} \mathcal{A}^{b \bar c}}{\bar z- \bar w} +... \end{align}
In section \ref{primaries} the concept of primary field will be defined precisely.
The coefficients appearing in the previous OPE will be explained, and we will compute the first subleading terms (see equation \eqref{jPhiO1}). 

Moreover, we propose that the following equations hold in the model under
consideration:
\be\label{dAdj} \partial \mathcal{A}^{a \bar a} =
-\frac{i{f^a}_{bc}}{c_+}:j^c_{L,z} \mathcal{A}^{b \bar a}: = -\frac{i{f^{\bar
      a}}_{\bar b \bar c}}{c_-}:j^{\bar c}_{R,z} \mathcal{A}^{a \bar b}: \ee
\be\label{dbarAdj} \bar \partial \mathcal{A}^{a \bar a} =
-\frac{i{f^a}_{bc}}{c_-}:j^c_{L,\bar z} \mathcal{A}^{b \bar a}:=
-\frac{i{f^{\bar a}}_{\bar b \bar c}}{c_+}:j^{\bar c}_{R,\bar z}
\mathcal{A}^{a \bar b}:. \ee 
One argument for the previous equations is the following. 
We start with the definition of the
adjoint operator in terms of the group element \eqref{PrimAdj}, and compute its
derivative:
\beq \p \mathcal{A}^{a \bar a} &=& x \, \p STr(g^{-1} t^a g t^{\bar a}) \cr
&=& x \, STr ( - g^{-1} \p g g^{-1} t^a g t^{\bar a} + g^{-1} t^a \p g g^{-1} g t^{\bar a} ) \cr
&=& x \, \frac{j^c_{L,z} \kappa_{dc}}{c_+} STr ( g^{-1}[t^a,t^d] g t^{\bar a} ) \cr
&=& -\frac{j^c_{L,z}}{c_+} i{f^a}_{bc} \mathcal{A}^{b \bar a}. \eeq
We have left out the normal ordering symbols from the above classical
calculation. The properties used in the calculation are that the
supertrace is graded cyclic and the fact that the equation
$g g^{-1}=1$ and its derivative hold true. We assume that the quantum theory is consistent with these two 
rules.
In section \ref{primaries} we will give a generic proof of equations \eqref{dAdj} and \eqref{dbarAdj},
 valid up to a certain order in a semi-classical expansion (see equation \eqref{dPhi=JPhi}).
 
Notice that the relations \eqref{dAdj} and \eqref{dbarAdj} imply that $\partial(\kappa_{ab} \mathcal{A}^{a
  \bar a} \mathcal{A}^{b \bar b})=0 =\bar \partial(\kappa_{ab} \mathcal{A}^{a \bar
  a} \mathcal{A}^{b \bar b})$ (and identical equations with the barred
indices contracted), and thus are compatible with the equations
relating the adjoint primary to the identity \eqref{AdjAdj=Ibar} and
\eqref{AdjAdj=I}.

\subsection{The left current - right current OPEs}
We have collected the tools to calculate the left/right current
operator product expansions.  Thanks to equations \eqref{jLAdj=jR} and
\eqref{jLAdjB=jRB} we only need the left current self OPEs
\eqref{euclidOPEs} as well as the OPE between the left current and the
adjoint primary operator \eqref{jAdj}.  As an example, we will explicitly
compute the OPE $j^a_{L,z}(z) j^{\bar a}_{R,z}(w)$ at the order of the
poles. We use the prescription of appendix \ref{compositeOPEs}:
\begin{align}\label{jLjR1stStep} j^a_{L,z}&(z)  j^{\bar a}_{R,z}(w) =  
-j^a_{L,z}(z) \frac{c_-}{c_+} \kappa_{cb}:j^b_{L,z} \mathcal{A}^{c \bar a}:(w) \cr
& = -\frac{c_-}{c_+} \kappa_{cb} \lim_{:x \to w:} 
\left[ j^a_{L,z}(z) j^b_{L,z}(x) \mathcal{A}^{c \bar a}(w) \right] \cr
& = -\frac{c_-}{c_+} \kappa_{cb} \lim_{:x \to w:} \left[
\left( \frac{c_1 \kappa^{ab}}{(z-x)^2} 
+ \frac{c_2 {f^{ab}}_d j^d_{L,z}(x)}{z-x} 
+ \frac{(c_2-g) {f^{ab}}_d j^d_{L,\bar z}(x)(\bar z - \bar x)}{(z-x)^2} 
+  ... \right) \mathcal{A}^{c \bar a}(w) \right. \cr
& \qquad + \left. j^b_{L,z}(x)
 \left( \frac{c_+}{c_++c_-} \frac{i{f^{ac}}_d 
\mathcal{A}^{d \bar a}(w)}{z-w} + ... \right) \right] \cr
& = -\frac{c_-}{c_+} \left[ \frac{c_1 \mathcal{A}^{a \bar a}(w)}{(z-w)^2} + \left(-c_2 + \frac{i c_+}{c_++c_-} \right) \frac{{f^a}_{bc}:j^b_{L,z}\mathcal{A}^{c \bar a}:(w)}{z-w} \right. \cr
& \qquad \left. - (c_2-g) \frac{{f^a}_{bc}:j^b_{L,\bar z}\mathcal{A}^{c \bar a}:(w)(\bar z - \bar w)}{(z-w)^2}  + ... \right]
\end{align}
In principle the second- and first-order poles that we obtain in the last line may receive corrections from the lower-order terms that we neglected in the penultimate line.
We will now argue that it is not the case.
Let us consider the first term in the last
line (the second-order pole). This term may receive corrections
of the form ${T^a}_b \mathcal{A}^{b \bar a}$, where the tensor ${T^a}_b$
contains at least one structure constant. Such a tensor vanishes by
using  properties of the Lie super algebras under consideration
\cite{Bershadsky:1999hk}.  Let us now consider the second
term (the holomorphic simple pole). It could  receive
corrections of the form ${T^a}_{bc}:j^b_{L,z}\mathcal{A}^{c \bar a}$, where
${T^a}_{bc}$ contains at least two structure constants. Again,
according to \cite{Bershadsky:1999hk}, this tensor vanishes because
traceless 
four-tensors invariantly contracted with structure constants over two indices
 vanish.
The third term receives no higher order corrections for the same reason.
Thus the terms written in the last line of \eqref{jLjR1stStep} are not corrected.
Using equations \eqref{dAdj} and \eqref{dbarAdj} we finally obtain:
\be\label{jLjR1} j^a_{L,z}(z) j^{\bar a}_{R,z}(w) = \frac{c_+ c_-}{c_++c_-}\left( \frac{\mathcal{A}^{a \bar a}(w)}{(z-w)^2}
+ \frac{c_-}{c_++c_-} \frac{\partial \mathcal{A}^{a \bar a}(w)}{z-w} + \frac{c_-}{c_++c_-} \frac{\bar \partial \mathcal{A}^{a \bar a}(w)(\bar z - \bar w)}{(z-w)^2} \right)  + ...
\ee
where the ellipses refer to terms of order zero or more in the distance between the two operators.
Similarly we can compute:
\begin{align} \label{jLjR2}
j^a_{L,\bar z}(z) j^{\bar a}_{R,\bar z}(w) &=  
\frac{c_+ c_-}{c_++c_-}\left( \frac{\mathcal{A}^{a \bar a}(w)}{(\bar z-\bar w)^2}
+ \frac{c_+}{c_++c_-} \frac{\partial \mathcal{A}^{a \bar a}(w)(z-w)}{(\bar z-\bar w)^2} + \frac{c_+}{c_++c_-} \frac{\bar \partial \mathcal{A}^{a \bar a}(w)}{\bar z-\bar w} \right)+ ... \cr
j^a_{L,z}(z) j^{\bar a}_{R,\bar z}(w) &= 
-\frac{c^2_+}{c_++c_-} \left(\mathcal{A}^{a \bar a}(w) 2\pi \delta^{(2)}(z-w)
- \frac{c_-}{c_++c_-} \frac{\partial \mathcal{A}^{a \bar a}(w)}{\bar z - \bar w} + \frac{c_-}{c_++c_-} \frac{\bar \partial \mathcal{A}^{a \bar a}(w)}{z - w}\right) + ...\cr
j^a_{L,\bar z}(z) j^{\bar a}_{R, z}(w) &= 
-\frac{c^2_-}{c_++c_-} \left(\mathcal{A}^{a \bar a}(w) 2\pi \delta^{(2)}(z-w)
+ \frac{c_+}{c_++c_-} \frac{\partial \mathcal{A}^{a \bar a}(w)}{\bar z - \bar w} - \frac{c_+}{c_++c_-} \frac{\bar \partial \mathcal{A}^{a \bar a}(w)}{z - w}\right)+ ...
\end{align}
The first two OPEs can be written in the alternative form:
\be j^a_{L,z}(z) j^{\bar a}_{R,z}(w) = \frac{c_+ c_-}{c_++c_-} 
\frac{\mathcal{A}^{a \bar a} (w)
\left( \frac{c_+ w + c_- z}{c_++c_-}\right)}{(z-w)^2}+ ...
\ee
\be j^a_{L,\bar z}(z) j^{\bar a}_{R,\bar z}(w) = 
\frac{c_+ c_-}{c_++c_-} \frac{\mathcal{A}^{a \bar a} (w) \left(\frac{c_- w + c_+ z}{c_++c_-}\right)}{(\bar z-\bar w)^2}+ ...
\ee
It is straightforward to show that the OPEs are compatible with
current conservation and the Maurer-Cartan equation.  These OPEs are
also compatible with the fact that the stress-tensor can be written
either in terms of the left-current or in terms of the right currents.
As an example of these consistency checks,
 it is shown in appendix \ref{TRJL} that when we express the energy-momentum 
tensor in terms of right currents, it satisfies the expected OPE with the left
current:
\be T(z) j^a_{L,z}(w) = \frac{1}{2c_3}\kappa_{\bar c \bar b}:j^{\bar b}_{R,z}j^{\bar c}_{R,z}:(z) j^a_{L,z}(w) = \frac{j^a_{L,z}(w)}{(z-w)^2} + \frac{\partial j^a_{L,z}(w)}{z-w} + \mathcal{O}\left((z-w)^0\right) \ee

When the theory is defined on a cylinder we can Fourier expand the
currents along the angular coordinate, at a given time. It was shown
in \cite{Ashok:2009xx} that the modes of the time components of the left (or
the right) currents generate an affine Lie algebra at level
$k$. The full commutator 
algebra computed  in appendix \ref{commutators} shows that these two
affine Lie algebras commute.

\subsection*{Summary}

In this section we have determined the pole order parts of the left
and right current operator product expansions. The
algebra closes on the current components and the adjoint field. Under
the assumptions on the quantum theory stated above, the coefficients
of the algebra are exact\footnote{If the assumptions are not valid,
  the coefficients will receive higher order corrections in
  $f^2$. The results in the rest of the paper are independent of these
possible corrections.}. We now move from the determination of the left-right
symmetry algebra of the model to the 
study of the vertex operators.

\section{The primaries}
\label{primaries}
In this section we 
define the concept of current algebra primaries.  These fields can be
understood as the elementary vertex operators of the conformal field
theory.  We compute the operator product expansion between a primary
field and a current perturbatively, and deduce the OPE between a
primary field and the stress-tensor.  In particular we derive the OPEs
used in \cite{Ashok:2009jw}.

\subsection*{Left current algebra primaries}
Given a representation $\mathcal{R}$ of the group $G_L$ we define a left primary field $\phi_\mathcal{R}$ with respect to
the left current algebra \eqref{euclidOPEs} as a field satisfying the operator product expansions:
\begin{align}  \label{defPrimaries}
 j_{L,z}^a(z,\bar z) \phi_\mathcal{R}(w,\bar w) &= - \frac{c_+}{c_+ + c_-} t^a \frac{\phi_\mathcal{R}(w,\bar w)}{z-w} + \text{order zero} \cr
 j_{L,\bar z}^a(z,\bar z) \phi_\mathcal{R}(w,\bar w) &=- \frac{c_-}{c_+ + c_-} t^a \frac{\phi_\mathcal{R}(w,\bar w)}{\bar z-\bar w} + 
\text{order zero} 
\end{align}
where the matrices $t^a$ are the generators of the Lie super-algebra
taken in the representation $\mathcal{R}$ associated to the primary
field $\phi_\mathcal{R}$.  If one assumes the above form for the
operator product expansions, then the coefficients of the poles are
fixed by the Ward identity for the symmetry $G_L$ and the demand that 
the contact term vanishes in the operator product expansion between 
the field $\phi$ and the
Maurer-Cartan operator \eqref{MC}. 
The Ward identity implies
compatibility of the OPEs \eqref{defPrimaries} with current
conservation \eqref{CC}.  An example of a left current primary is the
adjoint primary we discussed in the previous section.

In appendix \ref{WZWaffine} it is shown that 
a current primary field at a given point of the moduli space remains 
a current primary field after deformation of the kinetic term in the action. 
Thus one can consistently think of the current algebra primaries as 
the group element $g$ taken in the representation $\mathcal{R}$.
It also implies that at the WZW points the current primaries are 
the affine primary fields.

As argued in section \ref{bootstrap}, we can 
compute the less
singular terms in the current-primary OPE \eqref{defPrimaries} order
by order in $f^2$, by using the current conservation and the
Maurer-Cartan equation. 
Performing the calculation 
of higher order terms to order $f^2$, we find the OPE:
\begin{align}\label{jPhiO1}
j_{L,z}^a(z,\bar z) \phi(w,\bar w) = & \ - \frac{c_+}{c_+ + c_-} t^a \frac{\phi(w,\bar w)}{z-w} + :j^a_{L,z} \phi:(w,\bar w) \cr
& + {A^a}_c \log |z-w|^2 :j^c_{L,z}\phi:(w,\bar w) + {B^a}_c\frac{\bar z - \bar w}{z-w}:j^c_{L,\bar z}\phi:(w,\bar w) +... \cr
j_{L,\bar z}^a(z,\bar z) \phi(w,\bar w) = & \ - \frac{c_-}{c_+ + c_-} t^a \frac{\phi(w,\bar w)}{\bar z-\bar w} + :j^a_{L,\bar z} \phi:(w,\bar w) \cr
& - {A^a}_c \frac{z-w}{\bar z - \bar w} :j^c_{L,z}\phi: - {B^a}_c \log |z-w|^2 :j^c_{L,\bar z}\phi:(w,\bar w) +... 
\end{align}
where we dropped the subscript $\mathcal{R}$. The coefficients read:
\be\label{ABorder1} {A^a}_c =  \frac{c_-}{(c_++c_-)^2} i {f^a}_{cb} t^b + \mathcal{O}(f^4)
\qquad ; \qquad 
{B^a}_c =  \frac{c_+}{(c_++c_-)^2} i {f^a}_{cb} t^b + \mathcal{O}(f^4). \ee
The details of the calculation are given in appendix \ref{AppjPhi}. Note that
the coefficients of the simple poles are unmodified.

\subsection*{Current primaries are Virasoro primaries}
We will now show that a primary field with respect to the
left-current algebra is also a primary field with respect to the
Virasoro algebra.  The holomorphic worldsheet stress tensor is:
\be T(z) = \frac{1}{2c_1}\kappa_{ba}:j^a_{L,z} j^b_{L,z}:(z). \ee 
Let us consider
the OPE between a left-primary field $\phi$ and the holomorphic
stress-tensor:
\begin{align} 
\phi(z) 2 c_1 T(w) &= \lim_{:x \to w:}\phi(z) j^a_{L,z}(x) j^b_{L,z}(w) \kappa_{ba}.
\end{align}
{From} the structure of the OPE \eqref{jPhiO1}, and from the fact that
all operators appearing in this OPE are assumed to be composites of
currents and of the operator $\phi$, it follows that the most singular
term that may appear in this OPE is a double pole, multiplying the
operator $\phi$. As a consequence all the positive modes $L_{n>0}$ of
the holomorphic stress-tensor annihilate the operator $\phi$. Thus
this operator is a Virasoro primary.

Furthermore, with the knowledge of the current-primary OPE
\eqref{jPhiO1} up to order $f^2$, we can evaluate the stress-tensor/primary OPE up to the same order. Details about this computation are given in appendix \ref{AppTphi}. We obtain :
\begin{align}\label{Tphi}
T(w) \phi(z)
&= \frac{f^2}{2} \frac{t^a t^b \kappa_{ba} \phi(z)}{(z-w)^2} +\frac{1}{c_+} \frac{\kappa_{ba}t^a :j^b_{L,z}\phi:(z)}{w-z}+ \mathcal{O}(z-w)^{0}+ \mathcal{O}(f^4).
\end{align}
The same computation can be performed with the anti-holomorphic stress-tensor. We obtain:
\begin{align}\label{Tbarphi}
\bar T(\bar w) \phi(z)
&= \frac{f^2}{2} \frac{t^a t^b \kappa_{ba} \phi(z)}{(\bar z-\bar w)^2} +\frac{1}{c_-} \frac{\kappa_{ba}t^a :j^b_{L,\bar z}\phi:(z)}{\bar w-\bar z}+ \mathcal{O}(\bar z-\bar w)^{0}+ \mathcal{O}(f^4).
\end{align}
On general grounds the OPE between the stress-tensor and the primary
field $\phi$ reads:
\be T(w) \phi(z) = \frac{\Delta_{\phi} \phi(z)}{(w-z)^2} +
 \frac{\p \phi (z)}{w-z} + \mathcal{O}((z-w)^{0}), \ee
 where $\Delta_{\phi}$ is the left conformal dimension of the operator
 $\phi$.  Thus we deduce the conformal dimensions of the primary field
 $\phi$:
\be \Delta_{\phi}  = \bar \Delta_{\phi}  = \frac{f^2}{2} t^a t^b 
\kappa_{ba}+ \mathcal{O}(f^4).
\label{conformaldimension}
\ee 
The semi-classical result for the conformal dimension of a current
primary is as expected. It is equal to the quadratic Casimir of the
representation in which the field transforms, times the inverse radius
of the group manifold squared. 
For generic current primaries, there could be
corrections 
of order $f^4$ 
to this formula.  These corrections were conjectured to
be absent in \cite{Bershadsky:1999hk}. 
This was proven to be the case to
all orders in perturbation theory if the
superdimension of the representation of the primary is non-zero
(i.e. for short multiplets). For example for the short, 
discrete
representation crucial to the calculation in \cite{Ashok:2009jw}, there are no
corrections.

Notice that the stress-energy tensor
can also be written in terms of the right currents.  Equation
\eqref{conformaldimension} implies that a primary field transforms
under the left- and right-action of the group in representations that
have the same eigenvalue of the quadratic Casimir operator.
The simple poles in \eqref{Tphi} and \eqref{Tbarphi} also give the
relations:
\be\label{dPhi=JPhi} \p \phi(z) = \frac{1}{c_+} \kappa_{ba}t^a :j^b_{L,z}\phi:(z)+ \mathcal{O}(f^4) \ee
\be\label{dbarPhi=JbarPhi} \bar{\partial} \phi(z) = \frac{1}{c_-} \kappa_{ba}t^a :j^b_{L,\bar z}\phi:(z)+ \mathcal{O}(f^4). \ee

\subsubsection*{Remark about the atypical sector}

Some of the primary fields are associated to atypical Kac modules,
that are reducible but indecomposable \cite{Gotz:2006qp}.  In that
case the matrices $t^a$ that appear in equation \eqref{defPrimaries}
are not invertible.  Moreover the quadratic operator $\kappa_{ba}t^a
t^b$ can then be written in an upper-triangular form, with zeros on
the diagonal (which is the generalized eigenvalue of the quadratic
casimir for atypical representations of e.g. the $psl(n|n)$
superalgebra).  Equation \eqref{Tphi} tells us that the operator $L_0$
is proportional to this quadratic operator $\kappa_{ba}t^a t^b$ when
acting on a primary field. This implies that $L_0$ is
non-diagonalizable, which betrays the logarithmic nature of the theory
(see \cite{Gotz:2006qp} for a similar argument in the case of
$psl(2|2)$, and \cite{Gaberdiel:2001tr},\cite{Flohr:2001zs} for an
introduction to logarithmic CFTs).  Let us 
 remark here that the
fact that the current component $j_{L,z}$ has dimensions $(1,0)$, but
is not holomorphic also codes the logarithmic nature of the conformal
field theory \cite{Read:2001pz}.

\section{A recursive bootstrap 
for the elementary operator algebra 
}\label{bootstrap}
In this section we will explain how to compute
the current-current and current-primary OPEs
order by order 
in a semi-classical expansion. 
We will show that the knowledge of the poles in these OPEs
is enough to fix all the 
subleading terms. 
The idea driving the
bootstrap is to ask for the compatibility of the elementary
OPEs with both current conservation and the Maurer-Cartan equation.

\subsection*{Current-current OPEs}
First let us consider the current-current OPEs. Current conservation
gives the 
first 
constraints:
\beq j^a_{L,z}(z) \left[ \bar \p j^b_{L,z}(w) + \p j^b_{L,\bar z}(w)\right]
 = 0 \cr
j^a_{L,\bar z}(z) \left[ \bar \p j^b_{L,z}(w) + \p j^b_{L,\bar z}(w)\right] = 0. \eeq
The first line implies a one-to-one correspondence between the terms
in the $j^a_{L,z} j^b_{L,z}$ and $j^a_{L,z} j^b_{L,\bar z}$ 
OPEs. The second line then links the $j^a_{L,\bar z}
j^b_{L,\bar z}$ and the $j^a_{L,z} j^b_{L,\bar z}$ OPEs.  
These OPEs are expected to vanish up to contact terms. Indeed the same OPEs code the Ward identity for the global symmetry $G_L$. It follows that the contact terms in these OPEs are given by the transformation properties of the left current under the left action of the group on itself
\footnote{These contact terms allow for the computation of the holomorphic (respectively anti-holomorphic) poles in the $j^a_{L,z} j^b_{L,z}$ (respectively $j^a_{L,\bar z} j^b_{L,\bar z}$) OPE. These poles were already computed to all orders in $f^2$ in \cite{Ashok:2009xx} using different methods.}.

The second
constraint comes from the Maurer-Cartan equation :
\be j^a_{L,z}(z) \left[ c_- \bar \p j^b_{L,z}(w) - c_+ \p j^b_{L,\bar z}(w) + i {f^b}_{cd}:j^d_{L,z} j^c_{L,\bar z}:(w)\right] = 0. \ee
Contact terms in this OPE should vanish. 
Using current conservation and the fact that $c_++c_-=-f^{-2}$ we rewrite this constraint as :
\be\label{jmodMC} j^a_{L,z}(z) \bar \p j^b_{L,z}(w) = f^2 j^a_{L,z}(z) i {f^b}_{cd}:j^d_{L,z} j^c_{L,\bar z}:(w). \ee
Thanks to the factor of $f^2$ on the right-hand side of the previous
equation, it becomes manifest
 that the knowledge of the current algebra at a
given order in $f^2$ will also determine the current algebra at the
next order. The discussion of appendix \ref{XXOPEs} shows that the
terms appearing in the current-current OPEs at order $f^{2n}$ are
composites 
 of at most $n+1$ currents. This allows us to make an ansatz
for the current-current OPE at higher-order. Then equation
\eqref{jmodMC} fixes the coefficients in this ansatz. This method is
illustrated in appendix \ref{jMCOPE} where we compute the current
algebra up to order $f^2$.

\subsection*{Current-primary OPEs}
The same logic applies to 
 the computation of the current-primary
OPEs. Current conservation links the $j^a_{L,z} \phi$ and $j^a_{L,\bar
  z} \phi$ OPEs :
\be\label{phiCC} \phi(z) \left[ \bar \p j^a_{L,z}(w) + \p j^a_{L,\bar z}(w)\right] = 0. \ee
When the above equation is valid, the Maurer-Cartan constraint can be
rewritten as:
\be\label{phimodMC} \phi(z) \bar \p j^a_{L,z}(w) = f^2 \phi(z) i {f^a}_{cd}:j^d_{L,z} j^c_{L,\bar z}:(w). \ee
Again the discussion of appendix \ref{XXOPEs} gives an ansatz for the
current-primary OPE at a given order in $f^2$: the terms appearing in
the current-primary OPE at order $f^{2n}$ are composites
 of at most $n$
currents with the primary field $\phi$. When we plug this ansatz in
equation \eqref{phimodMC} we obtain the value of the
coefficients. This method is illustrated in appendix \ref{AppjPhi}
where we compute the current-primary OPE up to order $f^2$.

\subsection*{Further remarks}
This perturbative approach squares well with the observation
that the most singular terms in the current-current and
current-primary OPEs come with the lower power of $f^2$. This is
explained in appendix \ref{XXOPEs}. Thus performing a computation up
to a certain order in $f^2$ allows to truncate the current-current and
current-primary OPEs at a certain order in the distance between the
insertion points of the operators.

The consistency of this 
perturbative 
approach demands that the addition of
higher-order terms to the elementary OPEs does not spoil
their compatibility both with current conservation and with the
Maurer-Cartan equation at lower order in $f^2$.  That this is the case
is proven in appendix \ref{consistentPertOPEs}. 

One may hope to obtain a closed formula for the full
current-current and current-primary OPEs thanks to this
 algebraic
bootstrap.

\section{Composite operators and their conformal dimension}
\label{confdimcomp}
In this section we consider operators that are composites of one or more currents with a primary operator.
We are mostly interested in the computation of the conformal dimension
of such operators as a function of the two parameters $(k,f)$ of the 
supergroup sigma-model.  At the WZW point these operators
are descendants in the highest-weight
representations of the left affine Lie algebra.

\subsection*{Operators of the form $:j_L \phi :$}

Let us consider the operator $:j_{L,z}^a \phi:$ defined as the regular
term in the OPE between the operators $j_{L,z}^a$ and $\phi$.  To
compute the holomorphic dimension of this operator we compute its OPE
with the stress-tensor, and look at the second order pole. The
computation is done following the method described in appendix
\ref{compositeOPEs}. The fact that the stress-tensor is holomorphic
simplifies the calculation.  We find:
\begin{align} T(z) :j^a_{L,z}\phi:(w)
& = \lim_{:x \to w:} T(z) j^a_{L,z}(x) \phi(w) \cr
& = \lim_{:x \to w:} \left \{
\left( \frac{j^a_{L,z}(x)}{(z-x)^2} + \frac{\p j^a_{L,z}(x)}{z-x} \right) \phi(w) 
+ j^a_{L,z}(x) \left( \frac{\Delta_{\phi} \phi(w)}{(z-w)^2} + \frac{\p \phi(w)}{z-w} \right)
\right\} \cr
&= \lim_{:x \to w:} \left \{
\frac{1}{(z-x)^2}\left( 
-\frac{c_+}{c_++c_-} \frac{t^a \phi(w)}{x-w} + :j^a_{L,z} \phi:(w) \right.\right. \cr
&\qquad \qquad  \left. \left. 
+ {A^a}_c \log|x-w|^2 :j^c_{L,z} \phi:(w) + {B^a}_c \frac{\bar x - \bar w}{x-w} :j^c_{L,\bar z} \phi:(w) + ...
\right) \right. \cr
& \quad + \frac{1}{z-x}\left(
\frac{c_+}{c_++c_-} \frac{t^a \phi(w)}{(x-w)^2} + :\p j^a_{L,z} \phi:(w) \right. \cr
&\qquad \qquad  \left. \left. 
+ {A^a}_c \frac{1}{x-w} :j^c_{L,z} \phi:(w) - {B^a}_c \frac{\bar x - \bar w}{(x-w)^2} :j^c_{L,\bar z} \phi:(w) + ... 
\right) \right. \cr
& \quad \left. 
+ \frac{\Delta_{\phi}:j^a_{L,z} \phi:(w)}{(z-w)^2} + \frac{:j^a_{L,z}\p \phi:(w)}{z-w} 
\right\}  \cr
&=
-\frac{2}{(z-w)^3} \frac{c_+}{c_++c_-} t^a \phi(w) + \frac{:j^a_{L,z} \phi:(w)}{(z-w)^2}  \cr
&\qquad + \frac{1}{(z-w)^3} \frac{c_+}{c_++c_-} t^a \phi(w) + \frac{:\p j^a_{L,z} \phi:(w)}{z-w} + {A^a}_c \frac{1}{(z-w)^2} :j^c_{L,z} \phi:(w) \cr
& \quad + \frac{\Delta_{\phi}:j^a_{L,z} \phi:(w)}{(z-w)^2} + \frac{:j^a_{L,z}\p \phi:(w)}{z-w} + \mathcal{O}(z-w)^0
\end{align}
Using equation \eqref{ABorder1} we obtain :
\begin{align} \label{TjphiStep1}
T(z) :j^a_{L,z}\phi:(w) &= -\frac{c_+}{c_++c_-}\frac{t^a \phi(w)}{(z-w)^3}
+\frac{(\Delta_\phi +1):j^a_{L,z}\phi:(w) + \frac{c_-}{(c_++c_-)^2}i{f^a}_{cb}t^b:j^c_{L,z}\phi:(w)}{(z-w)^2} \cr
&\quad +\frac{\partial :j^a_{L,z}\phi:(w)}{z-w} + \mathcal{O}(f^4)+ \mathcal{O}(z-w)^0.
\end{align}
The matrices $t^a$ are the generators of the Lie algebra in the
representation in which the operator $\phi$ transforms. Since one has
a non-vanishing third-order pole, not all of the operators
$:j^a_{L,z}\phi:$ are Virasoro primary. Indeed we know 
from equation \eqref{dPhi=JPhi} 
that the
operator $L_{-1}\phi = \partial \phi$, which is a Virasoro descendant,
is a linear combination of these operators.  However the remaining
ones are all Virasoro primaries.  In the case where the quadratic
Casimir of the representation $\mathcal{R}$ associated to the
representation of the operator $\phi$ is non-zero, it is
straightforward to check that in the OPE between the stress-tensor and
the operator $ c^{(2)}_\mathcal{R} :j^a_{L,z}\phi: - t^a t_b
:j^b_{L,z}\phi: $, the third-order pole vanishes.  We adopt the
notation $c^{(2)}_{\mathcal{R}}$ both for the (generalized) quadratic
Casimir operator and for its eigenvalues in the irreducible
representation or reducible indecomposable structure $\mathcal{R}$.

{From} the double pole in equation \eqref{TjphiStep1} we can read off the
action of the scaling operator $L_0$ on the operator
$:j^a_{L,z}\phi:$:
\be L_0 :j^a_{L,z}\phi: = (\Delta_\phi +1):j^a_{L,z}\phi: 
+ \frac{c_-}{(c_++c_-)^2}i{f^a}_{cb}t^b:j^c_{L,z}\phi:.\ee
The operators $:j^a_{L,z}\phi:$ do not diagonalize the scaling operator
$L_0$.  In order to extract the conformal dimensions of these
operators we have to compute the eigenvalues of the following operator
:
\be\label{offDiagL0} {f^a}_{cb} {{(t^b)}_{\alpha}}^{ \beta} \ee
where we wrote explicitly the indices $\alpha$, $\beta$ associated to
the representation $\mathcal{R}$ in which the primary field $\phi$
transforms. This operator is an endomorphism acting on the vector
space associated to the tensor product of the adjoint and the
representation $\mathcal{R}$, namely $Adj \otimes \mathcal{R}$. Since
the structure constants are the generators of the Lie super-algebra in
the adjoint representation, the operator \eqref{offDiagL0} can be
rewritten as:
\be\kappa_{bd} t^d_{Adj} \otimes t^b_{\mathcal{R}}\ee
where this time we kept the external indices implicit.
The generators of the Lie super-algebra in the (reducible) representation $Adj \otimes \mathcal{R}$ read :
\be t^a_{Adj} \otimes Id + Id \otimes t^a_{\mathcal{R}}. \ee
Hence the quadratic Casimir operator in the tensor product of representations is :
\beq c^{(2)}_{Adj \otimes \mathcal{R}} &=&  \kappa_{ba} (t^a_{Adj} \otimes Id + Id \otimes t^a_{\mathcal{R}}) (t^b_{Adj} \otimes Id + Id \otimes t^b_{\mathcal{R}}) \cr
&=&  c^{(2)}_{\mathcal{R}} + c^{(2)}_{Adj} + 2 \kappa_{ba} t^a_{Adj} \otimes t^b_{\mathcal{R}}.\eeq
We deduce that the operator \eqref{offDiagL0} that we want to
diagonalize reads:
\be\kappa_{ba} t^a_{Adj} \otimes t^b_{\mathcal{R}} = \frac{1}{2} \left(
c^{(2)}_{Adj \otimes \mathcal{R}} - c^{(2)}_{\mathcal{R}} - c^{(2)}_{Adj}
\right). \ee
Recall that the quadratic Casimir vanishes in the adjoint
representation: $c^{(2)}_{Adj}=0$.  In the tensor product $Adj \otimes
\mathcal{R}_\phi$, reducible indecomposable structures may appear. The
Casimir operator is not diagonalizable on these structures, but we can
still define its generalized eigenvalues. 

Finally we obtain the conformal dimension of the operators $:j^a_{L,z} \phi:$.
 Let us denote by $\tilde{\mathcal{R}}$ a representation that
appears in the tensor product $Adj \otimes \mathcal{R}$, and by
$[ :j^a_{L,z} \phi: ]_{\tilde{\mathcal{R}}}$ a linear combination of the operators
$:j^a_{L,z} \phi:$ that transforms in the representation
$\tilde{\mathcal{R}}$. We have shown:
\begin{eqnarray}
h \left(\left[:j^a_{L,z} \phi:\right]_{\tilde{\mathcal{R}}} \right) &=& \frac{f^2}{2} c^{(2)}_\mathcal{R} + 1 + 
\frac{f^2}{2} (1-k f^2) (c^{(2)}_{\tilde{\mathcal{R}}}-c^{(2)}_\mathcal{R})+ \mathcal{O}(f^4).  
\end{eqnarray}
The interpretation of this semi-classical result is as follows. At
zero Wess-Zumino coupling $k=0$, we find that the conformal dimension
at leading order is $f^2 c^{(2)}_{\tilde{\mathcal{R}}}/2+1$, namely
the quadratic Casimir of the representation in which the total
wave-function $j \phi$ transforms times the inverse radius squared,
plus one for the fact that we are looking at a descendant state. That
is as for a naive evaluation of the conformal dimension of the
derivative of an ordinary point-particle wave function in
representation $\tilde{\mathcal{R}}$.  Note that at the WZW point $k
f^2 = 1$, we also recuperate the usual behavior, which is that only
the representation of the primary state $\phi$ counts for the basic
conformal dimension, while currents add precisely one to the conformal
dimension, independent of the representation in which the descendant
state transforms. Thus the formula interpolates between these two
intuitive behaviors, linearly in $kf^2$.  Notice that the corrections
to the dimension at the WZW point come from the logarithmic term in
the current-primary OPE \eqref{jPhiO1}.

This result illustrates the fact that the bulk
partition function will split into a sum over (mini-superspace)
representations of the supergroup with conformal dimensions depending
on the representation in question.
That demonstrates that this behavior, observed in boundary partition
functions \cite{Quella:2007sg}, extends to bulk partition functions.
This structure carries over to both left and right conformal
dimensions simultaneously.

Indeed, let us turn to the calculation of the anti-holomorphic
conformal dimension of the operator $:j^a_{L,z} \phi:$. As previously
we compute the OPE between the anti-holomorphic stress-tensor and the
operator:
\begin{align} \bar T(\bar z) &:j^a_{L,z}\phi:(w)
= \lim_{:x \to w:} \bar T(\bar z) j^a_{L,z}(x) \phi(w) \cr
& = \lim_{:x \to w:} \left \{
\left(  \frac{\bar \p j^a_{L,z}(x)}{\bar z-\bar x} \right) \phi(w) 
+ j^a_{L,z}(x) \left( \frac{\bar \Delta_{\phi} \phi(w)}{(\bar z-\bar w)^2} + \frac{\bar \p \phi(w)}{\bar z-\bar w} \right)
\right\} \cr
%
%
&=
\frac{:\bar \p j^a_{L,z} \phi:(w)}{\bar z-\bar w} + {A^a}_c \frac{1}{(\bar z-\bar w)^2} :j^c_{L,z} \phi:(w) \cr
& \quad + \frac{\bar \Delta_{\phi}:j^a_{L,z} \phi:(w)}{(\bar z-\bar w)^2} + \frac{:j^a_{L,z}\bar \p \phi:(w)}{\bar z-\bar w}
 \end{align}
Hence we have:
\begin{align} \bar T(\bar z) :j^a_{L,z}\phi:(w) &= \frac{\bar \Delta_{\phi}:j^a_{L,z}\phi:(w) - \frac{c_-}{(c_++c_-)^2}i{f^a}_{bc}t^b:j^c_{L,z}\phi:(w)}{(\bar z-\bar w)^2} + \frac{\bar \partial :j^a_{L,z}\phi:(w)}{\bar z-\bar w} + \mathcal{O}(f^4).
\end{align}
That leads to the conformal dimension:
\begin{eqnarray}
\bar h \left(\left[:j^a_{L,z} \phi:\right]_{\tilde{\mathcal{R}}} \right) &=& \frac{f^2}{2} c^{(2)}_\mathcal{R}  + 
\frac{f^2}{2} (1-k f^2) (c^{(2)}_{\tilde{\mathcal{R}}}-c^{(2)}_{\mathcal{R}})+ \mathcal{O}(f^4). \nonumber 
\end{eqnarray}
This is identical to the previous result, except for the lack of shift
by one (since we are acting with the holomorpic component of the left
current). 
Finally one can perform the same computation for the operators
$:j^a_{L,\bar z} \phi:$. One finds :
\begin{eqnarray}
h \left(\left[:j^a_{L,\bar z} \phi:\right]_{\tilde{\mathcal{R}}} \right) &=& \frac{f^2}{2} c^{(2)}_\mathcal{R}  + 
\frac{f^2}{2} (1+k f^2)( c^{(2)}_{\tilde{\mathcal{R}}}-c^{(2)}_{\mathcal{R}})+ \mathcal{O}(f^4)  \cr
\bar h \left(\left[:j^a_{L,\bar z} \phi:\right]_{\tilde{\mathcal{R}}} \right) &=& \frac{f^2}{2} c^{(2)}_\mathcal{R}  + 1 +
\frac{f^2}{2} (1+k f^2) (c^{(2)}_{\tilde{\mathcal{R}}}-c^{(2)}_{\mathcal{R}})+ \mathcal{O}(f^4) 
\end{eqnarray}

One can perform similar computations for operators that are composites
of a right-current and a primary operator. The conformal dimensions of
these operators can also be deduced from the symmetry of the model
under the simultaneous interchanges $g \leftrightarrow g^{-1}$ and $z
\leftrightarrow \bar z$.

\subsection*{A basis of operators}\label{spectrum}
At the Wess-Zumino-Witten points $k f^2 = 1$ the $\bar z$-component
(respectively $z$-component) of the left current (respectively right
current) vanishes, and the left current (respectively right current)
is holomorphic (respectively anti-holomorphic). Thus we can expand the
currents in a Laurent expansion.  The spectrum is generated by acting
on the affine primary fields with the modes of the current. It
is spanned by the operators :
\be \left \{ j^{a_1}_{-n_1}j^{a_2}_{-n_2}...j^{a_p}_{-n_p} \phi \right \} \ee
where $\phi$ is an affine primary operator and the currents $j^{a_i}$
can be either the left-current $J^{a_i}$ or the right-current $\bar
J^{a_i}$.  In fact, all the negative modes of the currents $J^a_{-n}$
(respectively $\bar J^a_{-n}$) can be generated by successive
commutations of the first negative mode $J^a_{-1}$ (respectively $\bar
J^a_{-1}$). This is most easily seen by working in the Chevalley basis
for the generators of the bosonic subalgebra. As a consequence the
spectrum is also spanned by the operators :
\be \left \{ j^{a_1}_{-1}j^{a_2}_{-1}...j^{a_p}_{-1} \phi \right \}. \ee
Finally, we notice that for any operator $\chi$, the operator
$J^a_{-1} \chi$ is the regular term in the OPE between the current
$J^a$ and the operator $\chi$. Thus we can rewrite $J^a_{-1} \chi =
:J^a \chi:$. So the previous set of operators spanning the spectrum
can be rewritten as:
\be \left \{ :j^{a_1}:j^{a_2}...:j^{a_p} \phi:...:: \right \}. \ee
We wrote the spectrum in this
 unusual form since it has the
advantage that these operators are also defined away from the WZW
point.

At a generic point of the moduli space both the left- and the
right-currents have two non-vanishing components. Since both left and
right invariant one-forms generate a basis for the cotangent bundle in
spacetime, the sets of operators generated by acting with
left-currents or with right-currents on primary fields are
isomorphic. This indicates that we have two (overcomplete) bases of
operators :
\be\label{proposalSpectrum} \left \{ :j^{a_1}_L:j^{a_2}_L...:j^{a_p}_L \phi:...:: \right \} = \left \{ :j^{a_1}_R:j^{a_2}_R...:j^{a_p}_R \phi:...:: \right \} \ee
where $\phi$ is a primary field as defined in section \ref{primaries},
and $j^{a_i}_L$ (respectively $j^{a_i}_R$) can be either the $z$- or
$\bar z$-component of the left current (respectively right current).
Of course, a mixture of left and right current components is also an
allowed choice.  We can compute the conformal dimensions of the
operators of the sets \eqref{proposalSpectrum} by following the
computation given at the beginning of this section.  The knowledge of
the current-current OPEs \eqref{euclidOPEs} and of the current-primary
OPEs \eqref{jPhiO1} up to terms of order $f^4$ allows the computation
of the conformal dimensions up to terms of order $f^4$.
 Following the
logic of section \ref{bootstrap} it is then possible to compute order
by order in $f^2$ the current-current OPEs, the current-primary OPEs
and finally the conformal dimensions of the operators
\eqref{proposalSpectrum}.  The recursive calculation may allow for a
closed solution.

Let us stress that the spectrum can be generated by acting with the
currents on a rather small set of primary operators. The current
primaries at any point of the moduli space are in one-to-one
correspondence with the affine primaries at the WZW points. In
particular the set of current primaries is much smaller than the set
of Virasoro primaries. Using the current algebra allows to take
advantage of the extension of the 
symmetry algebra at particular
points of the moduli space, namely the WZW points. In other words,
in the scheme proposed here, we
attempt to maximally exploit the presence of WZW lines in the
two-dimensional moduli space of $G_L \times G_R$ invariant supergroup
sigma-models.

\section{The classical and quantum integrability}\label{integrability}
The two-dimensional field theory under consideration is 
classically integrable in the sense that one can code the equations of
motion in the demand that a connection depending on a spectral
parameter is flat, thus leading to an infinite set of non-local
conserved charges. We give the proof of this fact for a generic
principal chiral model with Wess-Zumino term in appendix
\ref{classint}.

For the model to be quantum integrable, there needs to be an infinite
set of conserved charges in the quantum theory. There are
circumstances in which anomalies prevent the lifting of the charges
from the classical to the quantum theory. It is important to argue
that this is not the case for the supergroup sigma-models under
consideration here.

Beyond the usual conserved charges $Q^a_{(0)}$ 
associated to the group action(s) 
on itself, a first set of non-local conserved charges can be defined as
\cite{Luscher:1977uq}:
\begin{eqnarray}\label{Q1}
Q^a_{(1)} &=& N \int d \sigma j_\sigma^a + \int d \sigma_1 d \sigma_2
\epsilon(\sigma_1-\sigma_2) {f^a}_{bc} j^c_\tau(\tau,\sigma_1) j^b_\tau(\tau,\sigma_2),
\end{eqnarray}
where $\tau,\sigma$ are time- and space-coordinates, the factor $N$ is
an appropriate normalization constant, and the function $\epsilon$
takes the values $\pm 1$ depending on the sign of its argument. The
non-local charges exists
 for both left and right currents. The proof
of conservation of the non-local charge runs through the fact that the
current $j$ is conserved, and the validity of the Maurer-Cartan
equation.  When both equations are preserved in the quantum theory,
the (normal ordered) non-local charges survive (since from the first
non-local charges, all others can be generated through commutation
with the charges associated to the global symmetries).

It should be clear now that we can view the fixing of higher order
terms in the current-current operator products by demanding the
vanishing of Maurer-Cartan operator as demanding OPEs compatible with
the quantum integrability of the model. Conversely, the fact that one
can find such OPEs in this model (using the special algebraic
properties of the supergroup) lend credence to this hypothesis. It
would be useful to make the link between the existence of the Yangian
and the form of the current-current operator product expansions even
more manifest.

The main threat to the existence of the non-local charge 
\eqref{Q1} 
comes from
the UV-divergence in its definition. In the quadratic term, the
current components are both integrated, and the integration involves a
region in which the currents come very close 
to one another, thus
necessitating a UV regulator that could potentially render the
non-local charges anomalous. 

We will now show that in the models at hand, these potential UV
divergences are absent, at least to the first few orders in
perturbation theory, and presumably to all orders.  {From} the current
algebra \eqref{euclidOPEs}, we see that :
\beq {f^a}_{cb}\ j^b_{L}(z) j^c_{L}(w) &=& {f^a}_{cb}\ :j^b_{L}(z) j^c_{L}(w): + \mathcal{O}(f^4) \nonumber
\eeq
which is true for the $z$ and $\bar z$ components of the currents.
This follows from the fact that the tensors $\kappa^{bc}$,
${A^{bc}}_{de}$, ${B^{bc}}_{de}$ and ${C^{bc}}_{de}$ appearing in
\eqref{euclidOPEs} are graded-symmetric in the indices $b,c$.
Moreover the double contraction of structure constants (the Killing
form) also vanishes. 
 This is a proof of the consistency of the current
algebra with quantum integrability to
second order. It is a strong suggestion of quantum integrability to
all orders, a property which is closely tied to quantum conformal
invariance.

\section{Conclusions}\label{conclusions}

In this paper we continued the investigation of the conformal current
algebra in non-linear sigma models on supergroups. 
The left and right
current algebra closes on itself and a primary adjoint operator. 
The current algebra as well as the current-primary OPEs are tightly
constrained by the Maurer-Cartan equation and current conservation,
and can be computed order by order in a semi-classical expansion.
We argued that one can view the Hilbert space of the theory as
generated by currents acting on primaries, since WZW lines exist in
the moduli space of theory. We initiated the (perturbative)
computation of the spectrum, and argued for the possibility of a
recursive bootstrap.  We discussed the quantum integrability of the
model, and tied it to properties of the current algebra.
We hope our analysis contributes to the determination of an
explicit solution to the full bulk spectrum of two-dimensional conformal
field theories on supergroups and their cosets.

\section*{Acknowledgments}

We would like to thank Sujay Ashok, Costas Bachas, Denis Bernard, Vladimir Fateev,
Frank Ferrari,
Matthias Gaberdiel, Bernard
Julia, Anatoly Konechny, Thomas Quella, Sylvain Ribault and Walter Troost for useful
questions and helpful discussions.
J.T. would like to acknowledge support by ANR grant ANR-09-BLAN-0157-02.
The work of R.B. is supported in part by the Belgian Federal Science Policy Office through the Interuniversity Attraction Pole IAP VI/11 and by FWO-Vlaanderen through project G011410N.

\begin{appendix}

\section{Operator products involving composite operators}\label{compositeOPEs}

In this appendix we discuss the computation of OPEs involving composite
operators. We consider the following OPE:
\be\label{A:BC:} \lim_{z\to w} A(z) :BC:(w). \ee
The composite operator $:BC:(w)$ is defined as the term multiplied by $(x-w)^0
(\bar x - \bar w)^0$ in the OPE between the operators 
$B(x)$ and $C(w)$. To compute
the OPE \eqref{A:BC:} we use a point splitting procedure. We denote
the extraction of the normal ordered term by the limit
$:BC:(w) = \lim_{:x \to w:}B(x) C(w) $. This symbolizes that at the end
 of the calculation
we take the limit $x \to w$, and discard all terms that are
singular in $x-w$ in this limit.

To compute the operator product of the operator $A$ with the composite
operator $:BC:$ we proceed as follows. On the one hand we
perform the OPE of the operators $A$ and $B$, and then we perform the
OPE of the result with $C$. On the other hand we perform the OPE of
the operators $A$ and $C$, and then we perform the OPE of the result
with $B$. Eventually take the regular limit $:x \to w:$ and add up
the two terms.
Additional details about these operations follow.

\begin{itemize}
\item First let us consider the OPE between $A(z)$ and $B(x)$.  We
 evaluate the result at the point $x$ -- otherwise
taking the regular limit $:x \to w:$ would become cumbersome.  Let
  us consider one term in the OPE between $A(z)$ and $B(x)$:
\be\label{A:BC:step1} A(z) B(x)
 = ... + (z-x)^{\Delta_D - \Delta_A - \Delta_B}
(\bar z-\bar x)^{\bar\Delta_D - \bar\Delta_A - \bar \Delta_B} D(x)  + ...
\ee
where $\Delta_O$ (respectively $\bar \Delta_O$) stands for the
holomorphic (respectively anti-holomorphic) conformal dimension of an
operator $O$. For simplicity we consider a term in which no logarithm
appears, but the generalization is straightforward. We
have to perform the OPE of the right-hand side with the operator
$C(w)$. Let us consider one term in the result:
\begin{align}  &(z-x)^{\Delta_D - \Delta_A - \Delta_B}(\bar z-\bar x)^{\bar\Delta_D - \bar\Delta_A - \bar \Delta_B} D(x) C(w)
=\cr
&...+  (x-w)^{\Delta_E - \Delta_D - \Delta_C}(\bar x-\bar w)^{\bar\Delta_E - \bar\Delta_D - \bar \Delta_C}(z-x)^{\Delta_D - \Delta_A - \Delta_B}(\bar z-\bar x)^{\bar\Delta_D - \bar\Delta_A - \bar \Delta_B} E(w) +...
\nonumber
 \end{align}
Now to take the normal ordered limit $:x \to w:$, we expand the functions depending on $x$ in the neighborhood of $w$, namely, we write:
\be (z-x)^\Delta = (z-w)^\Delta -\Delta (x-w) (z-w)^{\Delta-1} + ... \ee
and we keep only the terms that end up with no factor of $(x-w)$.
The same manipulations have to be done
for the anti-holomorphic factors. If both
$\Delta_E - \Delta_D - \Delta_C$ and $\bar\Delta_E - \bar\Delta_D -
\bar \Delta_C$ are non-positive integers, then the term we isolated in
the previous steps contributes to the OPE \eqref{A:BC:} as:
\begin{align}\label{A:BC:step4}\lim_{z\to w}& A(z) :BC:(w) = ... + \# (z-w)^{\Delta_E - \Delta_A - \Delta_B-\Delta_C} (\bar z - \bar w)^{\bar\Delta_E - \bar\Delta_A - \bar\Delta_B-\bar\Delta_C} E(w)
\end{align}
with numerical coefficient:
\begin{align}\label{A:BC:step4coef}\# &= (-1)^{-\Delta_E + \Delta_D + \Delta_C}(-1)^{-\bar \Delta_E + \bar \Delta_D + \bar \Delta_C} \cr
& \times \frac{(\Delta_D - \Delta_A - \Delta_B)(\Delta_D - \Delta_A - \Delta_B-1)...(\Delta_E - \Delta_A - \Delta_B-\Delta_C+1)}{(-\Delta_E + \Delta_D + \Delta_C)!} \cr
& \times \frac{(\bar\Delta_D - \bar\Delta_A - \bar\Delta_B)(\bar\Delta_D - \bar\Delta_A - \bar\Delta_B-1)...(\bar\Delta_E - \bar\Delta_A - \bar\Delta_B-\bar\Delta_C+1)}{(-\bar\Delta_E + \bar\Delta_D + \bar\Delta_C)!} .
\end{align}
Let us stress that a given term in the result of the OPE \eqref{A:BC:}
may receive contributions from an $infinite$ number of terms in the
OPE between $A$ and $B$. This makes the computation of OPEs involving
composite operators rather involved.  One may need to resort to
perturbation theory in a small parameter to render the calculation
manageable. The perturbation theory that we use in the bulk of the
paper is explained in section \ref{bootstrap} and in the appendices \ref{XXOPEs} and \ref{consistentPertOPEs}.

\item Let us turn to the OPE between $A(z)$ and $C(w)$, which we
  evaluate at the point $w$. This second step is simpler than the
  first.  Again, we concentrate on one term in this OPE:
\be A(z) C(w) = ... + (z-w)^{\Delta_F - \Delta_A - \Delta_B}(\bar z-\bar w)^{\bar\Delta_F - \bar\Delta_A - \bar \Delta_B} F(w)  + ...\ee
We then have to perform the OPE between the right-hand side and the
operator $B(x)$. We evaluate the result at the point $w$. Let us write
down one term in the result:
\begin{align}  &(z-w)^{\Delta_F - \Delta_A - \Delta_B}(\bar z-\bar w)^{\bar\Delta_F - \bar\Delta_A - \bar \Delta_B} B(x) F(w)
=\cr
&...+  (z-w)^{\Delta_F - \Delta_A - \Delta_B}(\bar z-\bar w)^{\bar\Delta_F - \bar\Delta_A - \bar \Delta_B}(x-w)^{\Delta_G - \Delta_B - \Delta_F}(\bar x-\bar w)^{\bar\Delta_G - \bar\Delta_B - \bar \Delta_F} G(w) +... 
\nonumber
\end{align}
Finally we take the straightforward normal ordered limit $:x\to w:$, that discards all the terms except for the one with $\Delta_G - \Delta_B - \Delta_F=\bar\Delta_G - \bar\Delta_B - \bar \Delta_F=0$. 
Thus only the regular term $:BF:(w)$ in the OPE between $B(x)$ and $F(w)$ survives.  We obtain the following
contribution to the OPE \eqref{A:BC:}:
\begin{align}\lim_{z\to w}& A(z) :BC:(w) = ... + (z-w)^{\Delta_F + \Delta_B- \Delta_A -\Delta_C} (\bar z - \bar w)^{\bar\Delta_F+\bar \Delta_B - \bar\Delta_A -\bar\Delta_C} :BF:(w).
\nonumber
\end{align}

\end{itemize}

\subsection*{Simplification in the case of a holomorphic operator}

The computation of the singular terms in the OPE \eqref{A:BC:}
simplifies if the operator $A(z)$ is holomorphic.  Let us consider a
term of the form \eqref{A:BC:step1}. Since the operator $A$ is holomorphic there is
no dependence on $\bar z$, so $\bar\Delta_D - \bar\Delta_A - \bar
\Delta_B = 0$.  Let us also assume that $\Delta_D - \Delta_A -
\Delta_B$ is an integer.  The question is whether such a term may
contribute to a pole in the OPE \eqref{A:BC:}, i.e. a term of the form
\eqref{A:BC:step4} with $\Delta_E - \Delta_A - \Delta_B-\Delta_C$ a
negative integer (and $\bar\Delta_E - \bar\Delta_A -
\bar\Delta_B-\bar\Delta_C=0$).  But this is only possible if $\Delta_D
- \Delta_A - \Delta_B$ is already a negative integer, since otherwise the
coefficient \eqref{A:BC:step4coef} vanishes.

It follows from the previous discussion that under the assumption that
only integer powers of $(z-x)$ appear in the OPE between the operators $A(z)$ and
$B(x)$, then in the computation of singular terms in the OPE
\eqref{A:BC:} one can truncate the OPE between $A(z)$ and $B(x)$ to
the singular terms only (i.e. keep only the poles in $(z-x)$).
That specific feature of this special case is
 put to good use in some standard calculations in 
two-dimensional conformal field theory \cite{yellowbook}.

\section{The semi-classical behavior of the OPE coefficients}\label{XXOPEs}

At large radius, namely in the limit $f^2 \to 0$ (either at fixed
level $k$ or at fixed $kf^2$), the target space flattens and the
worldsheet theory becomes free. More precisely we obtain a theory of
$d$ free bosons, where $d$ is the dimension of the adjoint
representation of the super Lie algebra. Among these bosons, some are commuting
and some are anti-commuting, depending on whether they can be
associated to bosonic or fermionic coordinates of target space. At
fixed $kf^2$ the $f^2 \to 0$ limit is the semi-classical limit of the
model.

Our goal in this appendix is to evaluate the behavior at large radius
(small $f^2$) of the terms appearing in the current-current and
current-primary OPEs. Let us start with the action of the model:
\begin{align}
S &= S_{kin} + S_{WZ}\cr
S_{kin} &=  \frac{1}{ 16 \pi f^2}\int d^2 x Tr'[- \partial^\mu g^{-1}
\partial_\mu g]
\cr
S_{WZ} &= - \frac{ik}{24 \pi} \int_B d^3 y \epsilon^{\alpha \beta \gamma}
Tr' (g^{-1} \partial_\alpha g g^{-1} \partial_\beta g   g^{-1} \partial_\gamma g ).
\end{align}
We write the group element as:
\be g=e^{f X}=e^{i f X_a t^a} \ee
where the $X_a$ are coordinates on the supergroup and the matrices $t^a$ are the generators of the Lie superalgebra. 
The kinetic term and the Wess-Zumino term become:
\begin{eqnarray}\label{action(X)}
S_{kin} &=& \frac{1}{4 \pi}
\int d^2 z \left( \partial X_a \bar{\partial} X^a -
\frac{f^2}{12} {f^a}_{fe} {f}_{acb}  X^b \partial X^c X^e \bar{\partial} X^f
+ ...\right)
\nonumber \\
S_{WZ} &=&  -\frac{kf^2 }{12 \pi} 
\int d^2 z   \left( 
 f f_{abc} X^c  \partial X^b
  \bar{\partial} X^a + ... \right).
\end{eqnarray}
Written in this way the theory describes a set of interacting bosons
(some of which are anti-commuting).  The quadratic terms in the action
give rise to the free propagator:
\begin{eqnarray}\label{freeProp}
X^a(z,\bar{z}) X^b(w,\bar{w}) &=& - \kappa^{ab} \log \mu^2 |z-w|^2,
\end{eqnarray}
where $\mu$ is an infrared regulator. The propagator behaves like
$\mathcal{O}(f^0)$, whereas a vertex with $p+2$ legs (i.e. Lie algebra
indices) behaves as $\mathcal{O}(f^p)$.  It follows that the theory
reduces to a theory of free bosons in the semi-classical limit, as
anticipated.  At fixed $kf^2$ and for each interaction vertex, the
power of the coupling constant $f$ is equal to the number of structure
constants that appear.  Since we are interested in computing OPEs
involving the currents and the primary fields, let us write these
fields in terms of the bosons $X^a$:
\begin{eqnarray}\label{current(X)}
  \frac{j^a_{L,z}}{c_+} &=& (\partial g g^{-1})^a =  i (f \partial X^a + f^2 \frac{{f^{a}}_{bc}}{2} X^c \partial X^b + 
\frac{f^3}{6} {f^a}_{bc} {f^c}_{de}  \partial X^e X^d X^b +...)
  \nonumber \\
  \frac{j^a_{L,\bar z}}{c_-} &=& (\bar{\partial} g g^{-1})^a = i (f \bar{\partial} X^a + f^2 \frac{{f^{a}}_{cb}}{2} X^b \bar{\partial} X^c + 
\frac{f^3}{6} {f^a}_{bc} {f^c}_{de} \bar \partial X^e X^d X^b +...),
\end{eqnarray}
\be\label{phi(X)}
\phi = e^{i f X_a t^a} = i f X_a t^a - f^2 X_a t^a X_b t^b +... \ee
where in the last line
the generators $t^a$ are taken in the representation associated
to the primary field $\phi$.

\subsection*{The semi-classical behavior of the current-current OPE}
We study the semi-classical behavior of the OPE between two
$z$-components of the left-current.  The discussion generalizes
straightforwardly to other current-current OPEs.  We assume that the
only operators that appear in the result of this OPE are composites of
(derivatives of) left currents.  This is true at the WZW point, and
can presumably be proven at any point using conformal perturbation
theory.  Let us isolate one term in this OPE :
\be\label{jjOneTerm} j^a_{L,z}(z) j^b_{L,z}(w) 
= ... +  {A^{ab}}_{a_p a_{p-1}...a_{2} a_1}(z-w, \bar z - \bar w) :j^{a_1}_{L,z}:j^{a_2}_{L,z}...:j^{a_{p-1}}_{L,z}j^{a_p}_{L,z}:...::(w) +...\ee
Our goal is to evaluate the behavior of the tensor
${A^{ab}}_{a_p...a_1}(z-w, \bar z - \bar w)$ when the parameter $f$
is small. The reasoning will not depend on the particular current
component, nor on the presence of further derivative operators.

To proceed we use the expression \eqref{current(X)} of the currents in
terms of the bosonic fields $X^a$. First let us focus on the leading
term in the expansion \eqref{current(X)}. We consider the OPE:
\be\label{dXdXOneTerm} \p X^a(z) \p X^b(w) 
= ... +  {\tilde{A}^{ab}} {}_{a_p a_{p-1}...a_{2} a_1}(z-w, \bar z - \bar w) :\p X^{a_1}:\p X^{a_2}...:\p X^{a_{p-1}}\p X^{a_p}:...::(w) +...\ee
The behavior of the tensor ${\tilde{A}^{ab}} {}_{a_p a_{p-1}...a_{2}
  a_1}$ as a function of the parameter $f$ will give the behavior of
the tensor ${A^{ab}}_{a_p...a_1}(z-w, \bar z - \bar w)$ defined in
equation \eqref{jjOneTerm}.  As a first step let us consider the
following three-point function:
 \be\label{FeynmanDiag} \langle \p X^a(z) \p X^b(w) :\p X^{a_1}:\p X^{a_2}...:\p X^{a_{p-1}}\p X^{a_p}:...::(x) \rangle_{connected} \ee
 We consider only the contribution of connected Feynman diagrams to
 this correlation function. Indeed, if the external operators $\p
 X^a(z)$ and $\p X^b(w)$ are contracted on different pieces of a
 disconnected Feynman diagram, then the result contributes to the
 regular term $:\p X^a(x) \p X^b(w):$ on the right-hand side of the
 OPE \eqref{dXdXOneTerm}. Thus to compute the non-trivial terms in
 this OPE one needs to consider only the Feynman diagrams for which
 the external operators $\p X^a(z)$ and $\p X^b(w)$ are connected. But
 this in turn implies that the Feynman diagram is fully
 connected. Indeed, if this were not the case then one connected piece of the
 Feynman diagram has for external lines operators coming from the
 composite operator $ :\p X^{a_1}:\p X^{a_2}...:\p X^{a_{p-1}}\p
 X^{a_p}:...::(x)$ only. Such a piece would depend on the coordinate
 $x$ only, and would
  necessarily be zero by translation invariance.
This shows that we need to consider only fully
 connected Feynman diagrams.

 Now let us evaluate the $f$-dependence of a connected Feynman
 diagram contributing to \eqref{FeynmanDiag}. We will show by induction the following statement: a
 connected Feynman diagram in the theory \eqref{action(X)} with $p+2$
 external legs behaves like $\mathcal{O}(f^p)$. This is  the
 case for $p=0$ since the propagator is of order $f^0$. Now let us
 assume that the statement has been proven for $p <n+2$, and consider
 a Feynman diagram with $n+2$ external lines. We isolate $m$ of these
 external legs that are contracted on the same vertex with $m+1$
 legs. This piece is of order $f^{m-1}$. The other piece of the Feynman diagram has $n+2-m+1$ external lines, and by induction is of order $f^{n-m+1}$. Thus the result is of order
 $f^{n}$, and the proof is completed.
We deduce that:
 \be \langle \p X^a(z) \p X^b(w) :\p X^{a_1}:\p X^{a_2}...:\p X^{a_{p-1}}\p X^{a_p}:...::(x) \rangle_{connected} = \mathcal{O}(f^p).\ee
Since two-point functions of (composites of) the fields $X^a$
behave at least as $\mathcal{O}(f^0)$, we can now combine the previous result
with equation \eqref{current(X)} to evaluate the order
of the term in the current OPE under consideration\footnote{Using similar methods it can be shown that the subleading terms in equation
\eqref{current(X)} do not modify this conclusion}:
\be \frac{j^a_{L,z}(z)}{f c_+} \frac{j^b_{L,z}(w)}{f c_+} 
= ... +   \mathcal{O}(f^p) :\frac{j^{a_1}_{L,z}}{f c_+}:\frac{j^{a_2}_{L,z}}{f c_+}...:\frac{j^{a_{p-1}}_{L,z}}{f c_+}\frac{j^{a_p}_{L,z}}{f c_+}:...::(w) +...\ee
Given that $f c_+ = \mathcal{O}(f^{-1})$, we obtain:
\be j^a_{L,z}(z) j^b_{L,z}(w) 
= ... +  \mathcal{O}(f^{2p-2}) :j^{a_1}_{L,z}:j^{a_2}_{L,z}...:j^{a_{p-1}}_{L,z}j^{a_p}_{L,z}:...::(w) +...\ee
This is a property we repeatedly confirm as well as use in the bulk of
the paper.

\subsection*{The semi-classical behavior of the current-primary OPE}
We can perform a similar analysis to determine the behavior of
the terms in the current-primary OPE at large radius. Let us consider
a primary field $\phi$. We assume that all the terms that appear in
the OPE between a left current and this primary field are composite
operators including an arbitrary number of left currents and one field
$\phi$ only. This is the case at the WZW point. Then by continuously
deforming the OPEs away from the WZW point,  this is the
case over the whole moduli space of the theory.  Let us isolate one
term in the OPE between the left current $j^a_{L,z}$ and the primary
field $\phi$:
\be j^a_{L,z}(z) \phi(w) 
= ... +  {B^{a}}_{a_p a_{p-1}... a_1}(z-w, \bar z - \bar w) :j^{a_1}_{L,z}:j^{a_2}_{L,z}...:j^{a_p}_{L,z} \phi:...::(w) +...\ee
Our goal is to evaluate the behavior of the tensor
${B^{a}}_{a_p...a_1}(z-w, \bar z - \bar w)$ when the parameter $f^2$
is small. The composite operator we wrote down does not have any
derivative and contains only $z$-components of the left current, but
the result would be the same for a more general 
operator. Only the
number $p$ of currents will be relevant.  Following the previous
reasoning one can show that :
\be \langle \p X^a(z)  X^b(w) :\p X^{a_1}:\p X^{a_2}...:\p X^{a_{p}} X^{a_{p+1}}:...::(x) \rangle_{connected} = \mathcal{O}(f^{p+1}).\ee
Combining this result together with equations
\eqref{current(X)} and \eqref{phi(X)} and the fact that two-points functions are of order $\mathcal{O}(f^0)$ we get:
\be \frac{j^a_{L,z}(z)}{f c_+} \frac{\phi(w)}{if} 
= ... +  \mathcal{O}(f^{p+1}) :\frac{j^{a_1}_{L,z}}{f c_+}:\frac{j^{a_2}_{L,z}}{f c_+}...:\frac{j^{a_p}_{L,z}}{f c_+} \frac{\phi}{if}:...::(w) +...\ee
which we rewrite as:
\be j^a_{L,z}(z) \phi(w) 
= ... + \mathcal{O}(f^{2p}) :j^{a_1}_{L,z}:j^{a_2}_{L,z}...:j^{a_p}_{L,z} \phi:...::(w) +...\ee
This result on the order of magnitude of the operator product is
confirmed and used in the bulk of the paper.

\section{Consistency of 
perturbation theory}\label{consistentPertOPEs}

\subsection*{Current-current OPE}

In section \ref{bootstrap} we explained how to compute 
the current-current OPEs order by order in a semi-classical expansion. The idea is to ask for the vanishing of the OPE between a current and both current conservation and the Maurer-Cartan equation, order by order in $f^2$.
These two constraints can be combined as :
\be\label{AppjMC} j^a_{L}(z) \left(\bar \partial j^b_{L,z}(w)-i f^2 {f^b}_{cd}:j^d_{L,z} j^c_{L,\bar z}:(w)\right) = 0. \ee
For this 
perturbative 
method to be consistent a term of order 
$f^{2n}$ in the
current-current OPEs should not spoil the vanishing of the previous
OPE up to order $f^{2n-2}$.  The subtlety lies in the computation of
the OPE involving the composite operator in equation \eqref{AppjMC}.
Indeed the fact that the leading singularity in the current-current
OPE has a coefficient of order $f^{-2}$ threatens to generates terms
of low order in $f^2$ in this computation.  In this appendix we will
show that a term of order $f^{2n}$ in the current-current OPE does
produce terms of order $f^{2n}$ in the OPE between a current and the
composite operator appearing in equation \eqref{AppjMC}, namely $f^2
{f^b}_{cd}:j^d_{L,z} j^c_{L,\bar z}:$.

As a preliminary step let us prove the following useful lemma. We
consider a composite of $p$ currents $:j:j:j...j:...::$ that we write
symbolically $:j^p:$. Then the OPE of this operator with one current
$j$ is at most of order $f^{-2}$:
\be\label{lemmaf-2} j(z) :j^p:(w) = \mathcal{O}(f^{-2}). \ee
To prove this property we rewrite
the current in terms of the bosons $X^a$ using equation
\eqref{current(X)}. Schematically we have:
\be j = f^{-2} \sum_{n=0}^{\infty} \# f ^{n+1} :X^{n+1}: \ee
where we kept the numerical factors, possible derivatives acting on the fields $X$, and the index structures implicit to simplify the formula. Similarly the composite operator $:j^p:$ is written as:
\be :j^p: = f^{-2p} \sum_{m=0}^{\infty} \# f^{m+p} :X^{m+p}: \ee
To evaluate the OPE between the current $j$ and the composite operator $:j^p:$ 
we need to evaluate the OPE between operators of the form $:X^{q}:$. Remember that the propagator for the field $X$ is of order $f^0$, and that the $n$-point vertex is of order $f^{n-2}$. We deduce:
\be :X^{q_1}:(z) :X^{q_2}:(w) = \sum_{q=0}^{\infty} \mathcal{O}(f^{|q_1-q_2|-q}) :X^{q}:(z)  \ee
In the previous equation the estimation of the order of the terms is rough (especially for large $q$) but 
it
will be sufficient for our purposes.
The proof is similar to the argument given below \eqref{FeynmanDiag} (except that in the present case disconnected Feynman diagrams contribute).
We deduce an estimation for the order of the terms in the OPE \eqref{lemmaf-2}
\be\label{j.j^p} j(z) :j^p:(w) = f^{-2p-2} \sum_{n,m=0}^{\infty}  f^{n+m+p+1} \sum_{q=0}^{\infty} \mathcal{O}(f^{|n+1-m-p|-q}) :X^{q}: \ee
The operators that appear in the OPE \eqref{lemmaf-2} are themselves (composites of)
currents. Let us evaluate the coefficient of a composite operator of the form
$:j^r:$. According to equation \eqref{current(X)} the leading-order
term in this composite operator written in terms of $X$'s is :
\be :j^r: = f^{-r} :X^{r}: + \mathcal{O}(:X^{r+1}:). \ee
So to get the order of the coefficient that multiplies and operator $:j^r:$, 
it is enough to look for the coefficient of the terms multiplying
$f^{-r}:X^r:$ in the OPE \eqref{j.j^p}. These terms have a coefficient
of order: 
\be f^{-2p-2+n+m+p+1+|n+1-m-p|}=\left\{ \begin{array}{lll} 
f^{2(n+1-p)-2} & if & n+1 \ge m+p \\
f^{2m-2} & if & n+1 \le m+p.
 \end{array} \right. \ee
Thus this coefficient is of
order $\mathcal{O}(f^{-2})$. This completes the proof of \eqref{lemmaf-2}.

Now let us come back to the evaluation of the OPE between a current and the composite operator in equation \eqref{AppjMC}:
\be\label{modjMC3} j^a_{L,z}(z) i f^2 {f^b}_{cd}:j^d_{L,z} j^c_{L,\bar z}:(w) \ee
Let us
consider one term of order $f^{2n}$ in the OPE between the operators
$j^a_{L,z}$ and $j^d_{L,z}$, that we write schematically
$f^{2n}:j^p:$. To complete the computation we have to perform the OPE
of this operator with the remaining current $j^c_{L,\bar
  z}$. According to the previous lemma, this OPE produces terms with
coefficients of order $f^{-2}$. So we have proven that terms of order
$f^{2n}$ in the current-current OPE produce in the OPE \eqref{AppjMC} terms of order $f^{2 + 2n -
  2} = f^{2n}$.
This proves the consistency of the algorithm to compute the
current-current OPE order by order in $f^2$.

\subsection*{Current-primary OPE}

As explained in section \ref{bootstrap} the same logic allows us
to 
 perturbatively compute 
the operator product expansion between a current
and a primary operator.
The Maurer-Cartan equation can be combined with current conservation to give the constraint :
\be\label{phiModMC} \phi(z) \left(\bar \partial j^b_{L,z}(w)+i f^2 {f^b}_{cd}:j^d_{L,z} j^c_{L,\bar z}:(w)\right) = 0 \ee
This allows the computation of the $j^a_{L,z}.\phi$ OPE order by order in $f^2$.
The consistency of this algorithm is ensured by a slight generalization of lemma \eqref{lemmaf-2}, namely:
\be\label{lemmaf-2bis} j(z) :j^p \phi:(w) = \mathcal{O}(f^{-2}). \ee
The proof is similar to the proof of formula \eqref{lemmaf-2}.

\section{Conformal current algebra: precisions}\label{AppCurrents}
In this appendix we gather various technical results related to the current algebra \eqref{euclidOPEs}.

\subsection{The current algebra at order $f^2$}
\label{jMCOPE}
In \cite{Ashok:2009xx} the current algebra \eqref{euclidOPEs} was computed at the order of the poles. The discussion of section \ref{bootstrap} shows that we can compute the less-singular terms by demanding consistency with current conservation and the Maurer-Cartan equation. In this appendix we will give details of this computation, and derive in particular the value of the new coefficients \eqref{ABC} in the current algebra \eqref{euclidOPEs}.

In this
particular calculation, we show how to restore various signs that are
associated to the fact that we deal with a super Lie algebra. Since we 
use the special algebraic structure of supergroups with zero Killing form,
these signs are crucial.  To set up the problem,
we establish conventions for the metric inverse and the
contraction of indices:
\begin{eqnarray}
\kappa_{ab} \kappa^{cb} &=&  {\delta_a}^c
\nonumber \\
j_a &=& \kappa_{ab} j^b  \nonumber \\
{[} t_a, t_b {]} &=& i t_c {f^c}_{ab}.
\end{eqnarray}
We contract indices south-west north-east\footnote{These
conventions differ only slightly from those in \cite{Ashok:2009xx}.}.

As explained in section \ref{bootstrap} current conservation implies that the tensors $A,B,C$ that appear in each one of the three OPEs \eqref{euclidOPEs} are equal. To compute them  we ask for the vanishing of the OPE between a current and the Maurer-Cartan operator :
\begin{eqnarray}
c_- \partial_{\bar{z}} j_{L,z}^c 
- c_+ \partial_z j^c_{L,\bar z} - i {f^c}_{de} :j_{L,z}^e j_{L,\bar z}^d:.
\end{eqnarray}
Below we compute the OPE between the (left) current $j_{\bar z}^a$ and the Maurer-Cartan operator.
For ease of writing, we will separate various terms in the calculation.
We first calculate the operator product of the current with the first term:
\begin{eqnarray}
\mbox{Term 1} &=& j_{\bar z}^a (z) \cdot c_- \partial_{\bar{w}} j_z^c (w)
\nonumber \\
& \sim & c_- \partial_{\bar w} ( \tilde{c} \kappa^{ac} 2 \pi \delta(z-w)
\nonumber \\
& & 
+ {f^{ac}}_g (\frac{c_4-g}{\bar{z}-\bar{w}} j^g_z (z) + \frac{(c_2-g)}{z-w} j_{\bar{z}}^g (z)
\nonumber \\
& &
+ \frac{g}{4} \log |z-w|^2 (\partial_z j_{\bar{z}}^g(z) - \partial_{\bar z} j^g_z(z)))
\nonumber \\
& &
+ (-1)^{ac} : j_z^c j_{\bar z}^a :(z) 
\nonumber \\
& &
+ ( {(A)^{ac}}_{gh} \frac{\bar z - \bar w}{z-w} :j^{g}_{\bar z}
j^{h}_{\bar z}: (z) -  ( {(B)^{ac}}_{gh}  \log |z-w|^2  :j^{g}_{z}
j^{h}_{\bar z}: (z) \nonumber \\
& & +  ( {(C)^{ac}}_{gh}  \frac{z-w}{\bar{z}-\bar{w}} 
: j_z^{g} j_z^{h}:(z)))+... 
\end{eqnarray}
The second term we take into account comes from contracting the current with the
second term in the Maurer-Cartan operator:
\begin{eqnarray}
\mbox{Term 2} &=& j_{\bar z}^a (z) \cdot (-)c_+ \partial_{w} j_{\bar z}^c (w)
\nonumber \\
& \sim &- c_+ \partial_{w} (c_3 \kappa^{ac} \frac{1}{(\bar z - \bar w)^2}
\nonumber \\
& & 
+ {f^{ac}}_g 
(\frac{c_4}{\bar{z}-\bar{w}} j^g_{\bar z} (w) + \frac{(c_4-g)(z-w)}{(\bar z- \bar w)^2} 
j^g_{z} (w)
\nonumber \\
& &
+ \frac{g}{4} \frac{z-w}{\bar z - \bar w}
 (\partial_z j_{\bar{z}}^g(w) - \partial_{\bar z} j^g_z(w))
+\frac{c_4}{2} \partial_{\bar z} j^g_{\bar z}(w) 
+ \frac{c_4-g}{2} \frac{(z-w)^2}{(\bar z - \bar w)^2} \partial_z j^g_z (w))
\nonumber \\
& &
+ : j_{\bar z}^a j_{\bar z}^c :(w) 
\nonumber \\
& &
+ (-  {(A)^{ac}}_{gh} \log |z-w|^2 :j^g_{\bar z}
j^h_{\bar z}:+ {(B)^{ac}}_{gh} \frac{z-w}{\bar{z}-\bar{w}} :j^{g}_{z}
j^{h}_{\bar z}:
\nonumber \\
& &
+  {(C)^{ac}}_{gh} \frac{(z-w)^2}{(\bar{z}-\bar{w})^2} : j_z^g j_z^h:(w)))+...
\end{eqnarray}
Furthermore we have the contractions with the composite piece of the
Maurer-Cartan operator.
Following appendix \ref{compositeOPEs} we use a point-splitting procedure and write 
${f^c}_{de} :j_z^e j_{\bar z}^d:(w) = \lim_{:x \to w:} {f^c}_{de} j_z^e(x) j_{\bar z}^d(w)$.
Then we distinguish two terms. The simplest is the term where we contract
the current component $j_{\bar z}^a$ with the part at $w$ of the
split operator.  We then still need to contract further
while eliminating singularities as $x$ goes to $w$, but this is easily
done: only regular terms survive. We obtain:
\begin{eqnarray}
\mbox{Term 3} &=& (-i) (-1)^{ea}  {f^c}_{de} ( 
(c_3 \kappa^{ad} \frac{1}{(\bar z - \bar w})^2 j^e_z (w)
\nonumber \\
& & 
+ {f^{ad}}_g (\frac{c_4}{\bar{z}-\bar{w}} :j^e_z j^g_{\bar z}: (w) + 
\frac{(c_4-g)(z-w)}{(\bar z- \bar w)^2} 
:j^e_z j^g_{z}: (w) 
\nonumber \\
& & + \mbox{order zero in the separation.}
\end{eqnarray}
There is also the more involved  term where we contract first with
$j^e_z(x)$, and then further with $j^d_{\bar z}(w)$:
\begin{eqnarray}
\mbox{Term 4} &= & \lim_{:x \to w:} X^{ae}(z,x) (-i) {f^c}_{de}  j^d_{\bar z}(w)
\end{eqnarray}
where
\begin{eqnarray}
X^{ae}(z,x) 
& \sim & \tilde{c} \kappa^{ae} 2 \pi \delta(z-x)
\nonumber \\
& & 
+ {f^{ae}}_g (\frac{c_4-g}{\bar{z}-\bar{x}} j^g_z (z) + \frac{(c_2-g)}{z-x} j_{\bar{z}}^g (z)
\nonumber \\
& &
+ \frac{g}{4} \log |z-x|^2 (\partial_z j_{\bar{z}}^g(z) - \partial_{\bar z} j^g_z(z)))
\nonumber \\
& &
+ (-1)^{ae} : j_z^e j_{\bar z}^a :(z) 
\nonumber \\
& &
+   {A_{}^{ac}}_{gh} \frac{\bar z - \bar x}{z-x} :j^{g}_{\bar z}
j^{h}_{\bar z}: (z) -  {B_{}^{ac}}_{gh} \log |z-x|^2  :j^{g}_{z}
j^{h}_{\bar z}: (z)\nonumber \\
& &  +  {C_{}^{ac}}_{gh} \frac{z-x}{\bar{z}-\bar{x}}
 : j_z^{g} j_z^{h}:(z)
\nonumber \\
& &  + \mbox{order 1 in the separation and higher order in the parameter $f^2$.} \nonumber
\end{eqnarray}
Let's sum these four terms and discuss the vanishing of the total operator product order by
order.  The contact terms and double pole terms were already treated in 
\cite{Ashok:2009xx}. We cancel them as follows:

\noindent
1. There are terms proportional to $\partial_{\bar w} 2 \pi \delta(z-w)$.
These have coefficients:
\begin{eqnarray}
c_- \tilde{c} \kappa^{ac}
+c_+ c_3 \kappa^{ac} 
\end{eqnarray}
which vanishes since the coefficients of the current algebra \eqref{candg} satisfy :
\begin{eqnarray}
c_- \tilde{c} &=&- c_+ c_3
\end{eqnarray}

\noindent
2. There are terms proportional to $2 \pi \delta(z-w)$ with coefficient:
\begin{align}
- c_- & {f^{ac}}_g (c_2-g) j_{\bar z}^g (w)+ c_+ {f^{ac}}_g c_4 j_{\bar z}^g(w)
-i {f^c}_{de} \tilde{c} \kappa^{ae} j_{\bar z}^d (w)
\nonumber \\
& = - c_- {f^{ac}}_g (c_2-g) j_{\bar z}^g (w)+ c_+ {f^{ac}}_g c_4 j_{\bar z}^g(w)
-i (-1)^a (-1)^a {f^{ac}}_{g} \tilde{c}  j_{\bar z}^g (w)
\end{align}
which also vanishes thanks to the relation :
\begin{eqnarray}
-c_- (c_2-g) + c_+ c_4 - i \tilde{c} &=& 0.
\end{eqnarray}

\noindent
3. There are terms proportional to $1/(\bar z - \bar w)^2$ with
coefficients:
\begin{eqnarray}
c_-  (c_4-g) {f^{ac}}_g j^g_z + c_+ (c_4-g)  {f^{ac}}_g j_z^g
-i (-1)^{ea} {f^c}_{de} c_3 \kappa^{ad} j_z^e
-i {f^c}_{de} {f^{ae}}_g (c_4-g)^2 {f^{gd}}_h j^h_z
\nonumber
\end{eqnarray}
where the last term arises from expanding $1/(z-x)$ and taking
into account the further contraction in Term 4.
This last term vanishes thanks to the super-Jacobi identity combined with the vanishing of the Killing form.
Note that this implies that the second line in Term 4 does not contribute
when the contraction between $j^g_z$ and $j^d_{\bar z}$
gives rise to either a metric or structure constant. Thus, it can
potentially contribute starting at order zero in the separation only.
The coefficient of the terms under consideration then vanishes since
the coefficient \eqref{candg} satisfies the relation :
\begin{eqnarray}
(c_-+c_+)(c_4-g)  +i c_3 
&=& 0 .
\end{eqnarray}

\noindent
4. We now turn to the calculation which is new compared to \cite{Ashok:2009xx}.
In the operator product expansion the simple pole in $1/(\bar z - \bar w)$ comes with the coefficient :
\begin{align}
c_-&  {f^{ac}}_g (c_4-g) \partial_{\bar z} j_z^g(w)
-c_+ {f^{ac}}_g c_4 \partial_{z} j_{\bar z}^g (w)
\nonumber \\ &
-c_- \frac{g}{4} {f^{ac}}_g (\partial_z j_{\bar z}^g-
\partial_{\bar z} j_{ z}^g)
+c_-  {B^{ac}}_{gh}  : j_z^{g} j^{h}_{\bar z}:(w)
\nonumber \\ &
+c_+  \frac{g}{4} {f^{ac}}_g (\partial_z j_{\bar z}^g-
\partial_{\bar z} j_{ z}^g)
+c_+  {B^{ac}}_{gh}   : j_z^{g} j^{h}_{\bar z}:(w)
\nonumber \\ &
-i (-1)^{ea} {f^{c}}_{de} {f^{ad}}_g c_4 : j_z^e j_{\bar z}^g:(w)
\nonumber \\ &
-i  {f^{c}}_{de} {f^{ae}}_g (c_4-g) :j_z^g j^d_{\bar z}
\nonumber \\ &
 - i  {f^{c}}_{de} {f^{ae}}_g (c_2-g) {{B}^{gd}}_{xy} :j^{x}_{z} j^{y}_{\bar z}:
\nonumber \\ &
 +  \mathcal{O}(f^2) 
\end{align}
We use current conservation and the Maurer-Cartan equation to write:
\begin{align}
+i  &(c_4-\frac{g}{2}) {f^{ac}}_g
{f^g}_{de}  :j^e_z j^d_{\bar z}: 
+c_-  {B^{ac}}_{ed} : j_z^{e} j^{d}_{\bar z}:(w)
 \nonumber \\ &
+c_+ {B^{ac}}_{ed}   : j_z^{e} j^{d}_{\bar z}:(w)
 \nonumber \\ &
+(c_4-g/2) ( i {f^c}_{eg} {f^{ag}}_d (-1)^{ed} - i {f^c}_{dg} {f^{ag}}_e
)  : j_z^e j_{\bar z}^d:(w)
 \nonumber \\ &
+ g/2 (i {f^c}_{eg} {f^{ag}}_d (-1)^{ed} + i {f^c}_{dg} {f^{ag}}_e
): j_z^e j_{\bar z}^d:(w)
 \nonumber \\ &
- i  {f^{c}}_{hx} {f^{ax}}_g  (c_2-g) ){{B}^{gh}}_{ed} :j^{e}_{z} j^{d}_{\bar z}:
 \nonumber \\ &
+ \mathcal{O}(f^2)
\end{align}
where we have separated out (graded) symmetric and anti-symmetric
terms.  We now apply the super Jacobi identity to the first term in
the third line and note that:
\begin{eqnarray}
{f^{ce}}_{g} {f^{ag}}_d &=& {f^{ce}}_g {f^{ga}}_d (-1)^{1+a + ad}
\nonumber \\
&=& {f^{ec}}_g {f^{ga}}_d (-1)^{a + ad+ec}
\nonumber \\
&=& - (-1)^{a+ad+ec + cd}
( (-1)^{ac} { f^{ea}}_g {{f^g}_d}^c +
 (-1)^{ad} {f^e}_{dg} f^{gca}),
\end{eqnarray}
which leads to:
\begin{eqnarray}
  {f^c}_{eg} {f^{ag}}_d (-1)^{ed} -  {f^c}_{dg} {f^{ag}}_e
&=&(-1)^{1+a+ad+ec+cd+ed+ac+e+g+g+1+cd}
{f^c}_{dg} {f^{ag}_e}
\nonumber \\
& & + (-1)^{1+a+ad+ec+cd+ad+g+ed+g+g+ca+ed} {f^{ac}}_g {f^g}_{de}
\nonumber \\
& & 
- {f^c}_{dg} {f^{ag}}_e
\nonumber \\
&=& -  {f^{ac}}_g {f^g}_{de}.
\end{eqnarray}
Therefore, the third line cancels the first term in the first line
and we are left with:
\begin{align}
(c_- &  {(B)^{ac}}_{ed}+ c_+
 {(B)^{ac}}_{ed}) : j_z^{e} j^{d}_{\bar z}:(w)
 \nonumber \\ &
+ g/2 (i {f^c}_{eg} {f^{ag}}_d (-1)^{ed} + i {f^c}_{dg} {f^{ag}}_e
)
 : j_z^e j_{\bar z}^d:(w)
 \nonumber \\ &
- i  {f^{c}}_{hx} {f^{ax}}_g (c_2-g){{B}^{gh}}_{ed} :j^{e}_{z} j^{d}_{\bar z}: + \mathcal{O}(f^2).
\label{final}
\end{align}
As expected the demand of the vanishing of this term gives the value of the tensor $B$ at the first non-trivial order in $f^2$ :
\begin{eqnarray}
  {B^{ac}}_{ed}   &=& -i \frac{g}{2(c_++c_-)}
({f^c}_{eg} {f^{ag}}_d (-1)^{ed} +  {f^c}_{dg} {f^{ag}}_e) + O(f^4). 
\end{eqnarray}

\noindent 
5. A similar analysis for the other two first-order poles proportional respectively to 
 $1/( z -  w)$ and  $( z -  w)/(\bar z - \bar w)^2$
 gives respectively the tensors $A$ and $C$ in equations (\ref{euclidOPEs}, \ref{ABC}).
 The details of the calculation are very similar to the
calculation we just discussed.

\subsection*{Remarks on higher order terms in $f^2$}
To discuss a few aspects of the higher order terms that we encountered,
it is useful to define the following tensor:
\begin{eqnarray}
{S^{ac}}_{ed} &=& 
 {f^c}_{eg} {f^{ag}}_d (-1)^{ed} +  {f^c}_{dg} {f^{ag}}_e.
\end{eqnarray}
It is manifestly graded symmetric in the lower indices. Let's also check that
it is graded symmetric in the upper indices:
\begin{eqnarray}
{S^{ca}}_{ed} &=& 
 {f^a}_{eg} {f^{cg}}_d (-1)^{ed} +  {f^a}_{dg} {f^{cg}}_e
\nonumber \\
 &=& {f^{ag}}_e (-1)^{g+eg+1+ed}
 {f^c}_{dg} (-1)^{gd+1}
+ (-1)^{g+1+gd+1+eg}  {f^c}_{eg} {f^{ag}}_d
\nonumber \\
 &=& {f^{ag}}_e (-1)^{ac}
 {f^c}_{dg} 
+ (-1)^{ac}  {f^c}_{eg} {f^{ag}}_d
\nonumber \\
&=& (-1)^{ac} {S^{ac}}_{ed}.
\end{eqnarray}
Therefore, $S$ is a linear operator that acts
on the 
space of (graded) symmetric two-tensors.

The higher order term in the last line in the above explicit calculation \eqref{final}
gives rise to the square of the linear operator $S$. We computed it for
$psl(2|2)$ for which it simplifies to
\begin{eqnarray}
{S^{ac}}_{gh} {S^{gh}}_{ed} &=& 8 ( \kappa^{ac} \kappa_{de}
+ ( \delta^a_e \delta^c_d + (-1)^{ed} \delta^a_d \delta^c_e)).
\end{eqnarray}
We also have the equality $S^3=16S$.
When we take a supertrace of $S^2$, it can be shown to be zero because
the superdimension of
 $psl(2|2)$ is $-2$.

Using some of these properties, it is clear that at higher order 
the structure of a pole in the
$j^a \cdot MC^c$ OPE will look like:
\begin{eqnarray}
\dots \kappa^{ac} :j_{e\bar z} j_{z}^e :
+ 
\dots (:j^a_{\bar z} j^c_{ z} :
+
(-1)^{ac} :j^c_{\bar z} j^a_{ z} :).
\end{eqnarray}
The first term is proportional to a component of the energy-momentum
tensor
 (and to the kinetic term in the Lagrangian).  
The other term indicates that at higher order, we need a new
four-tensor index structure in the current-current operator product
expansion. At the same time, the special properties of the linear
operator given above show that only few four-tensors will appear.  It
is certainly feasible to push the above calculation, and therefore the other
calculations in the bulk of the paper to higher order.

\subsection{The Virasoro algebra from the current algebra}\label{TjandTT}
In \cite{Ashok:2009xx} it was shown that the Virasoro algebra emerges
from the current algebra \eqref{euclidOPEs} via the Sugawara
construction.  More precisely it was argued that the normal ordered
classical expression for the stress tensor :
\be T = \frac{1}{2 c_1} :j_{L,zb} j^{b}_{L,z}: \ee
satisfies the OPEs :
\be\label{TjApp} T(z) j^a_{L,z}(w) = \frac{j^a_{L,z}(w)}{(z-w)^2} +  \frac{\p j^a_{L,z}(w)}{z-w} + \mathcal{O}(z-w)^0 \ee
\be\label{TTApp} T(z) T(w) = \frac{sdim(G)}{2(z-w)^4} + \frac{T(w)}{(z-w)^2} +  \frac{\p T(w)}{z-w} + \mathcal{O}(z-w)^0. \ee

In this section, we fill a gap in the demonstration of equation
\eqref{TjApp}.  We reconsider the OPE between a current and the
bilinear operator $:j_{L,zb} j^{b}_{L,z}:$.  To perform this
computation in \cite{Ashok:2009xx} we truncated the current algebra at
the order of the poles.  We obtained :
\beq\label{j:jj:} j_{L,z}^a(z) :j_{L,zb} j^b_{L,z}:(w) &=& 2c_1
\frac{j_{L,z}^a(w)}{(z-w)^2} + c_2 \frac{ {f^a}_{bc}}{z-w}\left( (-1)^{bc} :j^b_{L,z}
  j^c_{L,z}:+ :j^c_{L,z} j^b_{L,z}:(w)\right) \cr &&+
   (c_2-g) {f^a}_{bc}
\frac{\bar z - \bar w}{(z-w)^2}\left( (-1)^{bc}:j^b_{L,z} j^c_{L,\bar z}:+:
  j^c_{L,\bar z} j^b_{L,z}:(w) \right) \nonumber \\
& &   +... \eeq
where the ellipses contain terms of order zero in the distance between
the insertion points $z$ and $w$.  We will now show that the
subleading terms in the current algebra do not modify this result.
Let us divide these terms into two sets. First we have the regular
terms and the terms that multiply an $n^{th}$-derivative of a single
current. These terms were already considered in \cite{Ashok:2009xx}
and it is straightforward to show that they do not modify
\eqref{j:jj:}.  The second set contains the terms that multiply
composites of (derivatives of) several currents (not including the
regular terms). This includes for instance the current bilinears in
equation \eqref{euclidOPEs}. The crucial point is that all these terms
come with a coefficient that contains at least two structure
constants. This is a consequence of the discussion in appendix
\ref{XXOPEs}.  In full generality, a term in this second set may lead
to the following type of contribution to \eqref{j:jj:}:
\begin{align} \label{j:jj:Mod} 
& \frac{{T^a}_b  j_{L,z}^b(w)}{(z-w)^2}
+\frac{{U^a}_b \p j_{L,z}^b(w)}{z-w} + \frac{{V^a}_b \bar \p j_{L,z}^b(w)(\bar z - \bar w)}{(z-w)^2}+\frac{{\bar T^a}_b j_{L,\bar z}^b(w)(\bar z - \bar w)}{(z-w)^3} \cr
& 
+\frac{{\bar U^a}_b \p j_{L,\bar z}^b(w)(\bar z - \bar w)}{(z-w)^2} + \frac{{\bar V^a}_b \bar \p j_{L,\bar z}^b(w)(\bar z - \bar w)^2}{(z-w)^3}+\frac{ {W^a}_{bc}:j^c_{L,z} j^b_{L,z}:(w) }{z-w}\cr
& 
+\frac{{X^a}_{bc}:j^c_{L,\bar z} j^b_{L,z}:(w)(\bar z - \bar w)}{(z-w)^2}
+\frac{{Y^a}_{bc}:j^c_{L,\bar z} j^b_{L,\bar z}:(w)(\bar z - \bar w)^2}{(z-w)^3} \end{align}
where the tensors ${T^a}_b$, etc. are invariant two- and three-tensors
made of contractions of structure constants.  According to the
argument of \cite{Bershadsky:1999hk}, any invariant two-tensor
obtained by contracting at least one structure constant
vanishes. Moreover any invariant three-tensor obtained by contracting
at least two structure constant also vanishes. Since all tensors
appearing in \eqref{j:jj:Mod} contain at least two structure constants
that come from the current-current OPE, all these terms vanish.  This
completes the proof of equation \eqref{TjApp}.

\subsection{Currents as a primary fields of dimension one revisited}\label{TRJL}
The stress-energy tensor can be written either in terms of the left or
of the right currents.  As a consistency check on our formalism, we
will compute in this appendix the OPE between the stress-energy tensor
and the current components $j_{L,z}$ using the expression of the
energy-momentum tensor $T$ in terms of the right currents:
\be T(z) = \frac{1}{2c_3} :j^{\bar b}_{R,z}j^{\bar c}_{R,z}:(w)\kappa_{\bar c \bar b} \ee
To proceed we use the OPEs between left and right currents
\eqref{jLjR1}, \eqref{jLjR2}, as well as the OPEs between a current
and the primary adjoint operator \eqref{jAdj}. Notice that 
the latter OPE may 
receive higher-order corrections in $f^2$. In the following we keep
track only of the leading-order terms in $f^2$. The computation goes
as follows:
\begin{align} j^a_{L,z}(z)& :j^{\bar b}_{R,z}j^{\bar c}_{R,z}:(w)\kappa_{\bar c \bar b}
= \frac{c_+ c_-}{c_++c_-} \left( \frac{[:\phi^{a \bar b}j^{\bar c}_{R,z}:(w) + :j^{\bar b}_{R,z} \phi^{a \bar c}:(w)]\kappa_{\bar c \bar b}}{(z-w)^2} \right. \cr
&\qquad + \frac{c_-}{c_++c_-}\frac{[:\partial \phi^{a \bar b}j^{\bar c}_{R,z}:(w) + :j^{\bar b}_{R,z} \partial \phi^{a \bar c}:(w)]\kappa_{\bar c \bar b}}{z-w} \cr
& \left.\qquad  + \frac{c_-}{c_++c_-}\frac{[:\bar \partial \phi^{a \bar b}j^{\bar c}_{R,z}:(w) + :j^{\bar b}_{R,z} \bar \partial \phi^{a \bar c}:(w)]\kappa_{\bar c \bar b}(\bar z-\bar w)}{(z-w)^2}
\right)
\end{align}
where a triple pole vanishes since it is proportional to the contraction of a structure constant with the metric.
We have to treat carefully the normal-ordered operators appearing in the previous expression. The central
point is the property:
\be\label{jphi-phij} :j^{\bar a}_R \phi^{b \bar b}: - :\phi^{b \bar b} j^{\bar a}_R: \propto {f^{\bar a \bar b}}_{\bar c}\ee
and similarly for the left currents. This property
 follows from the OPE between the current and the adjoint primary \eqref{jAdj}.
Thus we can deal with the first line
easily, and using equation \eqref{jRAdj=jL} we obtain :
\be \frac{[:\phi^{a \bar b}j^{\bar c}_{R,z}:(w) + :j^{\bar b}_{R,z} \phi^{a \bar c}:(w)]\kappa_{\bar b \bar c}}{(z-w)^2}
= -\frac{c_-}{c_+} \frac{2 j^a_{L,z}(w)}{(z-w)^2} .\ee 
Now let us consider the second line. We use equation \eqref{dAdj}:
\be \partial \phi^{a \bar a} = -\frac{i {f^{\bar a}}_{\bar b \bar c}}{c_-} j^{\bar c}_{R,z} \phi^{a \bar b} .\ee
Notice that we do not need the normal ordering symbol on the
right-hand side since (at leading order in $f^2$) there is no singular term to discard in the OPEs \eqref{jAdj}.
Thus we
rewrite the second line as:
\begin{align} \frac{c_-}{c_++c_-} & \frac{[ :\partial \phi^{a \bar b} j^{\bar c}_{R,z}: (w)  + :j^{\bar b}_{R,z} \partial \phi^{a \bar c}:(w)]\kappa_{\bar b \bar c}}{z-w} \cr
&= \frac{-i}{c_++c_-} \frac{[:{f^{\bar b}}_{\bar d \bar e} j^{\bar e}_{R,z} \phi^{a \bar d} j^{\bar c}_{R,z}:(w)  + :j^{\bar b}_{R,z} {f^{\bar c}}_{\bar d \bar e} j^{\bar e}_{R,z} \phi^{a \bar d}:(w)]\kappa_{\bar b \bar c}}{z-w} \cr
& = \frac{-i f_{\bar c \bar d \bar e}}{c_++c_-} \frac{:j^{\bar e}_{R,z} j^{\bar c}_{R,z} \phi^{a \bar d}:(w)  + :j^{\bar c}_{R,z} j^{\bar e}_{R,z} \phi^{a \bar d}:(w)}{z-w}
\end{align}
where we used the property \eqref{jphi-phij} again in the last step to
commute the adjoint operator and the current in the normal ordered
triple operator. Now thanks to the anti-symmetry of the structure
constants this term vanishes.  We can perform similar manipulations on
the third line:
\begin{align}
\frac{c_-}{c_++c_-} & \frac{[:\bar \partial \phi^{a \bar b}j^{\bar c}_{R,z}:(w) + :j^{\bar b}_{R,z} \bar \partial \phi^{a \bar c}:(w)]\kappa_{\bar b \bar c}(\bar z-\bar w)}{(z-w)^2}.
\end{align}
The first operator can be rewritten as:
\be :\bar \partial \phi^{a \bar b}j^{\bar c}_{R,z}:(w)\kappa_{\bar b \bar c} = 
-\frac{i f_{\bar c \bar d \bar e}}{c_+} : j^{\bar e}_{R,\bar z} j^{\bar c}_{R,z} \phi^{a \bar d}:(w)
\ee
and the second one as:
\be :j^{\bar b}_{R,z} \bar \partial \phi^{a \bar c}:(w)\kappa_{\bar b \bar c} =
-\frac{i f_{\bar c \bar d \bar e}}{c_+} : j^{\bar c}_{R,z} j^{\bar e}_{R,\bar z} \phi^{a \bar d}:(w) = :\bar \partial \phi^{a \bar b}j^{\bar c}_{R,z}:(w)\kappa_{\bar b \bar c}
\ee
where in the last step we used that $f_{\bar c \bar d \bar e}: j^{\bar
  c}_{R,z} j^{\bar e}_{R,\bar z}:= f_{\bar c \bar d \bar e}:j^{\bar
  e}_{R,\bar z}j^{\bar c}_{R,z} :$. Therefore the two operators
present on the third line are the same.  Now we can use the Maurer-Cartan
equation and current conservation for the right currents to rewrite
them as:
\be 
-\frac{i f_{\bar c \bar d \bar e}}{c_+} : j^{\bar e}_{R,\bar z} j^{\bar c}_{R,z} \phi^{a \bar d}:(w)
= - \frac{c_++c_-}{c_+}:\bar \partial j^{\bar b}_{R,z} \phi^{a \bar d}:(w)\kappa_{\bar b \bar d}.
\ee
Therefore, we can rewrite the third line as:
\be 
\frac{c_-}{c_++c_-}\frac{2 (c_++c_-)}{c_-} \frac{[: \bar \partial j^{\bar b}_{R,z} \phi^{a \bar c}:(w)+:j^{\bar b}_{R,z} \bar \partial \phi^{a \bar c}:(w)]\kappa_{\bar b \bar c}(\bar z-\bar w)}{(z-w)^2} 
= -2 \frac{c_+}{c_-}\frac{\bar \partial j^a_{L,z}(w) (\bar z-\bar w)}{(z-w)^2} 
\ee
where we used equation \eqref{jRAdj=jL} once more.
Gathering all terms, we obtain:
\be j^a_{L,z}(z) :j^{\bar b}_{R,z}j^{\bar c}_{R,z}:(w)\kappa_{\bar b \bar c} = 
-\frac{2c_-^2}{c_++c_-}\left( \frac{j^a_{L,z}(w)}{(z-w)^2} + \frac{\bar \partial j^a_{L,z}(w)(\bar z - \bar w)}{(z-w)^2} \right)\ee
which we can finally rewrite in the expected form:
\be T(w) j^a_{L,z}(z) = \frac{1}{2c_3} :j^{\bar b}_{R,z}j^{\bar c}_{R,z}:(w)\kappa_{\bar b \bar c} j^a_{L,z}(z)
 =  \frac{j^a_{L,z}(w)}{(w-z)^2} + \frac{\partial j^a_{L,z}(w)}{w-z},\ee
 thus completing our consistency check.

\subsection{The associativity of the current algebra}\label{associativity}

In this appendix we address the issue of the associativity of the current algebra \eqref{euclidOPEs}. We will prove the associativity of this current algebra at the first non-trivial order in $f^2$.

\subsection*{The OPE $j^a_{L,z}(z) j^b_{L,z}(w) j^c_{L,z}(x)$}

First we consider the OPE between three $z$-components of the left-current:
\be j^a_{L,z}(z) j^b_{L,z}(w) j^c_{L,z}(x). \ee
We will compute this OPE using the current algebra \eqref{euclidOPEs}
at the order of the poles. Moreover we will only compute the
lowest-order terms in the $f^2$ expansion. In this case these are
terms of order $f^{-2}$. To prove associativity we will first compute
the OPE between the first two currents, then compute the OPE of the
result with the third current, and show that the result is invariant
under permutation of the currents. We start out with:
\begin{align}
j^a_{L,z}(z) & j^b_{L,z}(w) j^c_{L,z}(x) = 
\left(
\frac{c_1 \kappa^{ab}}{(z-w)^2} + \frac{c_2 {f^{ab}}_d j^d_{L,z}(w)}{z-w}+ \frac{(c_2-g) {f^{ab}}_d j^d_{L,\bar z}(w)(\bar z - \bar w)}{(z-w)^2} \right. \cr
& \left. \qquad 
+ :j^a_{L,z}(z) j^b_{L,z}(w): + ...
\right) j^c_{L,z}(x).
\end{align}
The ellipses stand for lower-order terms in the OPEs, that we do not keep track of. We obtain :
\begin{align}
j^a_{L,z}(z) & j^b_{L,z}(w) j^c_{L,z}(x) = 
\frac{c_1 \kappa^{ab}j^c_{L,z}(x)}{(z-w)^2} + \frac{c_1 c_2 {f^{abc}}}{(z-w)(w-x)^2} + :j^a_{L,z}(z) j^b_{L,z}(w):j^c_{L,z}(x) + ...
\end{align}
up to a contact terms. We now have to compute the OPE involving the regular operator $:j^a_{L,z}(z) j^b_{L,z}(w):$. In order to use the techniques presented in appendix \ref{compositeOPEs} we rewrite both currents as being evaluated at the point $w$:
\be :j^a_{L,z}(z) j^b_{L,z}(w): = \sum_{n,\bar n=0}^{\infty} \frac{(z-w)^n }{n ! }\frac{ (\bar z - \bar w)^{\bar n}}{ \bar n !} :(\p^n \bar \p ^{\bar n} j^a_{L,z}) j^b_{L,z}:(w). \ee
Let us now consider the OPE of one of these composite operators with the current  $j^c_{L,z}(x)$:
\begin{align}
 j^c_{L,z}(x) & :(\p^n \bar \p ^{\bar n} j^a_{L,z}) j^b_{L,z}:(w) =
 j^c_{L,z}(x) \lim_{:y \to w:} \p_y^n \bar \p_y ^{\bar n} j^a_{L,z}(y) j^b_{L,z}(w) \cr
& = \lim_{:y \to w:} \p_y^n \bar \p_y ^{\bar n} \left[ 
\left ( \frac{c_1 \kappa^{ca}}{(x-y)^2} + \frac{c_2 {f^{ca}}_d j^d_{L,z}(y)}{x-y}+ \frac{(c_2-g) {f^{ca}}_d j^d_{L,\bar z}(y)(\bar x - \bar y)}{(x-y)^2} \right.\right. \cr
& \left. \qquad \qquad 
+ \sum_{m,\bar m=0}^{\infty} \frac{(x-y)^m }{m ! }\frac{ (\bar x - \bar y)^{\bar m}}{ \bar m !} :(\p^m \bar \p ^{\bar m} j^c_{L,z}) j^a_{L,z}:(y) + \mathcal{O}(f^2)
 \right) j^b_{L,z}(w) \cr
 & \left. \qquad + j^a_{L,z}(y) \left ( \frac{c_1 \kappa^{cb}}{(x-w)^2} + ... \right )
 \right] \cr
 & = \lim_{:y \to w:} \p_y^n \bar \p_y ^{\bar n} \left[ 
 \frac{c_1 \kappa^{ca}j^b_{L,z}(w)}{(x-y)^2} + \frac{c_1 c_2 {f^{cab}} }{(x-y)(y-w)^2} + \frac{c_1 \kappa^{cb}j^a_{L,z}(y)}{(x-w)^2}
 +.. \right]
 \end{align}
 where the ellipses in the last line contains singular terms that
 comes from the OPE between the regular operators and the current in
 the third line of the previous computation. These terms in this OPE
 will be removed by the regular limit $:y \to w:$. In order to compute
 the action of the derivatives more conveniently, we rewrite the
 second term in the last line as:
\be \frac{c_1 c_2 {f^{cab}} }{(x-y)(y-w)^2} = c_1 c_2 {f^{cab}} \sum_{p=0}^{\infty} \frac{(y-w)^{p-2}}{(x-w)^{p+1}} \ee
Thus we obtain:
\begin{align}
 j^c_{L,z}&(x)  :(\p^n \bar \p ^{\bar n} j^a_{L,z}) j^b_{L,z}:(w) =
 \lim_{:y \to w:}  \left[
 \delta_{\bar n,0}\frac{(n+1)!}{(x-y)^{n+2}} c_1 \kappa^{ca}j^b_{L,z}(w) \right. \cr
& \left. \qquad + \delta_{\bar n,0}\sum_{p=0}^{\infty} \frac{(p-2)...(p-2-n+1)(y-w)^{p-2-n}}{(x-w)^{p+1}} c_1 c_2 {f^{cab}}
 + \frac{c_1 \kappa^{cb} \p^n \bar \p ^{\bar n} j^a_{L,z}(y)}{(x-w)^2}+...
 \right] \cr
 & = 
 \delta_{\bar n,0}\frac{(n+1)!}{(x-w)^{n+2}} c_1 \kappa^{ca}j^b_{L,z}(w)
  + \delta_{\bar n,0} \frac{n!}{(x-w)^{n+3}} c_1 c_2 {f^{cab}}
  + \frac{c_1 \kappa^{cb} \p^n \bar \p ^{\bar n} j^a_{L,z}(w)}{(x-w)^2}+ ...
 \end{align}
Resumming the series, we get:
\begin{align}
:j^a_{L,z}(z) & j^b_{L,z}(w): j^c_{L,z}(x) = \sum_{n,\bar n=0}^{\infty} \frac{(z-w)^n }{n ! }\frac{ (\bar z - \bar w)^{\bar n}}{ \bar n !} 
\left[ \delta_{\bar n,0}\frac{(n+1)!}{(x-w)^{n+2}} c_1 \kappa^{ca}j^b_{L,z}(w) \right. \cr
& \qquad \left.
  + \delta_{\bar n,0} \frac{n!}{(x-w)^{n+3}} c_1 c_2 {f^{cab}}
  + \frac{c_1 \kappa^{cb} \p^n \bar \p ^{\bar n} j^a_{L,z}(w)}{(x-w)^2}+ ... \right] \cr
& = \frac{c_1 \kappa^{ca}j^b_{L,z}(w)}{(x-z)^2}
+ \frac{c_1 c_2 {f^{cab}}}{(x-z)(x-w)^2}
+ \frac{c_1 \kappa^{cb} j^a_{L,z}(z)}{(x-w)^2}
+...
\end{align}
After gathering all terms, we obtain:
\begin{align}
j^a_{L,z}(z) & j^b_{L,z}(w) j^c_{L,z}(x) = 
\frac{c_1 \kappa^{ab}j^c_{L,z}(x)}{(z-w)^2} + \frac{c_1 c_2 {f^{abc}}}{(z-w)(w-x)^2} +  \frac{c_1 \kappa^{ca}j^b_{L,z}(w)}{(x-z)^2} \cr
& \qquad + \frac{c_1 c_2 {f^{cab}}}{(x-z)(x-w)^2}
+ \frac{c_1 \kappa^{cb} j^a_{L,z}(z)}{(x-w)^2}
+... \cr
& = \frac{c_1 c_2 {f^{abc}}}{(z-x)(x-w)(w-z)} + \frac{c_1 \kappa^{ab}j^c_{L,z}(x)}{(z-w)^2}+  \frac{c_1 \kappa^{ca}j^b_{L,z}(w)}{(x-z)^2}
 +\frac{c_1 \kappa^{cb} j^a_{L,z}(z)}{(x-w)^2}+ \mathcal{O}(f^0) +... \nonumber
\end{align}
which is manifestly invariant under permutation of the currents.

\subsection*{The OPE $j^a_{L,z}(z) j^b_{L,z}(w) j^c_{L,\bar z}(x)$}

We now consider the OPE involving two $z$-components and one $\bar z$-component of the left current:
\be j^a_{L,z}(z) j^b_{L,z}(w) j^c_{L,\bar z}(x). \ee
First we will take first the OPE between the two $z$-components of the current:
\begin{align}
[ j^a_{L,z}(z) & j^b_{L,z}(w) ] j^c_{L,\bar z}(x) = 
\left(
\frac{c_1 \kappa^{ab}}{(z-w)^2} + \frac{c_2 {f^{ab}}_d j^d_{L,z}(w)}{z-w}+ \frac{(c_2-g) {f^{ab}}_d j^d_{L,\bar z}(w)(\bar z - \bar w)}{(z-w)^2} \right. \cr
& \left. \qquad 
+ :j^a_{L,z}(z) j^b_{L,z}(w): + ...
\right) j^c_{L,\bar z}(x) \cr
& = \frac{c_1 \kappa^{ab}j^c_{L,\bar z}(x)}{(z-w)^2} + \frac{c_3(c_2-g) {f^{abc}} (\bar z - \bar w)}{(z-w)^2(\bar w - \bar x)^2} + :j^a_{L,z}(z) j^b_{L,z}(w):j^c_{L,\bar z}(x) + ...
\end{align}
The OPE involving the composite operator does not produces any term of order $f^{-2}$, thus we obtain :
\be [ j^a_{L,z}(z) j^b_{L,z}(w) ] j^c_{L,\bar z}(x) = \frac{c_1 \kappa^{ab}j^c_{L,\bar z}(x)}{(z-w)^2} + \frac{c_3(c_2-g) {f^{abc}} (\bar z - \bar w)}{(z-w)^2(\bar w - \bar x)^2} + \mathcal{O}(f^0) +... \ee

Now let us perform the same computation taking first the OPE between one $z$-component and one $\bar z$-component of the current:
\begin{align}
 j^a_{L,z}(z) & [j^b_{L,z}(w)  j^c_{L,\bar z}(x)] \cr & = 
 j^a_{L,z}(z) 
 \left( \frac{(c_4-g) {f^{bc}}_d j^d_{L,z}(x)}{\bar w-\bar x}+ \frac{(c_2-g) {f^{bc}}_d j^d_{L,\bar z}(x)}{(w-x)}+ :j^b_{L,z}(w)  j^c_{L,\bar z}(x): +... \right) \cr
& = \frac{c_1(c_4-g) {f^{abc}}}{(z-w)^2(\bar w - \bar x)} + j^a_{L,z}(z):j^b_{L,z}(w)  j^c_{L,\bar z}(x): +... \cr
& = \frac{c_1(c_4-g) {f^{abc}}}{(z-w)^2(\bar w - \bar x)} + \frac{c_1 \kappa^{ab}j^c_{L,\bar z}(x)}{(z-w)^2} +...
\end{align}
Thanks to the relations between the coefficients of the current algebra :
\be c_1(c_4-g) = c_3(c_2-g) \ee
we find that the current algebra is indeed associative at the order at which we performed the computation. The coordinate dependence does not match exactly since we did not take into account the terms containing derivatives of the currents that appear in the current algebra as subleading terms. It is interesting
to pursue the full proof of associativity.

\subsection{The holomorphy of the stress-tensor}

In this appendix we address the issue of the holomorphy of the stress-tensor\footnote{We would like
to thank Matthias Gaberdiel for raising the issue.}:
\be T(z) = \frac{1}{2 c_1} \kappa_{ab} :j_{L,z}^b j_{L,z}^a:(z). \ee
Since the $z$-component of the left-current is not holomorphic away from the WZW point, it is not obvious that the stress-tensor will be holomorphic in the quantum theory.
The anti-holomorphic derivative of the stress-tensor reads:
\be \bar \p T(z) = \frac{1}{2 c_1} \kappa_{ab} \left( :\bar \p j_{L,z}^b j_{L,z}^a:(z) + :j_{L,z}^b \bar \p j_{L,z}^a:(z) \right). \ee
To continue the computation we combine current conservation 
with the Maurer-Cartan equation
to write the anti-holomorphic derivative of the $z$-component of the current in terms of a bilinear :
\be \bar \p j_{L,z}^a = -i f^2 {f^a}_{bc} :j^c_{L,z} j^b_{L,\bar z}:. \ee
Since all the poles in the OPE between $j^c_{L,z}$ and $ j^b_{L,\bar z}$ vanish when contracted with the structure constant ${f^a}_{bc}$, we can also write :
\be \bar \p j_{L,z}^a = -i f^2 {f^a}_{bc} :j^b_{L,\bar z} j^c_{L,z}:. \ee
Thus using successively the last two equations we obtain:
\be \bar \p T(z) = \frac{-i f^2}{2 c_1} f_{abc} \left( : :j^c_{L,z} j^b_{L,\bar z}: j_{L,z}^a:(z) + :j_{L,z}^a  :j^b_{L,\bar z} j^c_{L,z} ::(z) \right). \ee
Now let us consider the composite operator $::j^c_{L,z} j^b_{L,\bar z}: j_{L,z}^a:(z)$. It is defined as the regular term in the OPE between $
:j^c_{L,z} j^b_{L,\bar z}:$ and $j_{L,z}^a$. 
We will show that we have :
\be\label{f::::=f::} f_{abc}::j^c_{L,z} j^b_{L,\bar z}: j_{L,z}^a:(z) = f_{abc}:j^c_{L,z} j^b_{L,\bar z} j_{L,z}^a:(z) \ee
where the operator $:j^c_{L,z} j^b_{L,\bar z} j_{L,z}^a:$ is defined as the regular term in the OPE of the three currents $j^c_{L,z}$, $j^b_{L,\bar z}$ and $j_{L,z}^a$.
The difference between the operators on the left-hand side and the right-hand side of equation \eqref{f::::=f::} comes from the non-regular terms in the OPE between $j^c_{L,z}$ and $j^b_{L,\bar z}$. 
The crucial point is that all these terms vanish when contracted with the structure constant $f_{abc}$:
\be f_{abc} [j^c_{L,z}(z) j^b_{L,\bar z}(w) - :j^c_{L,z}(z) j^b_{L,\bar z}(w):] = 0. \ee
This can be checked via the current algebra OPEs \eqref{euclidOPEs}
order by order in $f^2$. In equation \eqref{euclidOPEs} the current
algebra is given up to terms of order $f^4$, and thus one can prove
the previous statement up to terms of order $f^4$. Indeed, all tensors that appear in the current algebra
\eqref{euclidOPEs} vanish upon double contraction with a structure
constant:
\be f_{abc} \left[ \kappa^{cb},\ f^{cbd},\ {A^{cb}}_{de},\ {B^{cb}}_{de},\ {C^{cb}}_{de} \right] = 0 \ee
The non-degenerate metric $\kappa^{cb}$ and the tensors
${A^{cb}}_{de},\ {B^{cb}}_{de},\ {C^{cb}}_{de}$ are graded-symmetric
in the indices $c,b$. Moreover the double contraction of the structure
constant vanishes since the dual Coxeter number of the Lie
super algebra vanishes. This concludes the proof of equation
\eqref{f::::=f::} up to terms of order $f^4$. 
Let us mention that the same equation \eqref{f::::=f::} also guaranties the quantum integrability of the model 
up to this order, as discussed in section \ref{integrability}.
The same argument leads to the equality:
\be f_{abc} :j_{L,z}^a  :j^b_{L,\bar z} j^c_{L,z} ::(z) 
= f_{abc} :j_{L,z}^a  j^b_{L,\bar z} j^c_{L,z} :(z). \ee
Thus we have:
\be \bar \p T(z) = \frac{-i f^2}{2 c_1} f_{abc} 
\left( : j^c_{L,z} j^b_{L,\bar z} j_{L,z}^a:(z) + :j_{L,z}^a  j^b_{L,\bar z} j^c_{L,z} :(z) \right) =0 \ee
which vanishes thanks to the 
(graded) anti-symmetry of the structure constants. 
It would be interesting
to have a non-perturbative understanding of the consistency of the normal-ordering
and the holomorphy of the
energy-momentum tensor.

\section{Details on primary operators}\label{AppPrimaries}

\subsection{Behavior of current primaries under perturbation of the kinetic term}\label{WZWaffine}

In this section we will show that current primaries at a given point
of moduli space remain current primaries after perturbation of the
kinetic term. More precisely we will show that if an operator $\phi$
satisfies the OPEs \eqref{defPrimaries} at a given point of moduli
space, then it also satisfies the same OPEs after exactly marginal
deformation of the theory.  This implies that it is consistent to
think of a current primary as being the group element taken in a given
representation, at any point of the moduli space.  It also proves the
claim in section \ref{primaries} that the affine primary fields at the
WZW points become current primaries after deformation of the theory.

For convenience let us recall the OPEs that define a primary operator $\phi$ :
\begin{align} \label{defPrimariesBis}  
 j^a_{L,z}(z) \phi(w) &= - \frac{c_+}{c_++c_-} t^a \frac{\phi(w)}{z-w} + \text{less singular} \cr
 j^a_{L,\bar z}(z) \phi(w) &= - \frac{c_-}{c_++c_-} t^a \frac{\phi(w)}{\bar z-\bar w} + \text{less singular.} 
\end{align}
We assume that these OPEs hold at a given point of moduli space
$(f^2,k)$. Then we perturb the kinetic term : $f^2 \to f^2 + \epsilon$
and we compute the way the OPEs \eqref{defPrimariesBis} are modified.
A procedure to compute OPEs in conformal perturbation theory was given
in \cite{Ashok:2009xx}. Here we will only compute the deformation of
the OPEs \eqref{defPrimariesBis} up to first order in $\epsilon$. The
prescription is to compute first the OPE between the current and the
perturbation of the action, and then to compute the OPE of the result
with the field $\phi$.  We begin with the first step of this
procedure, for the first OPE in \eqref{defPrimariesBis}. The OPE
between the current and the marginal operator can be computed thanks
to the current algebra \eqref{euclidOPEs}:
\begin{align}\label{jMarg} j^a_{L,z}(z)& \frac{\epsilon}{4\pi f^4} \int d^2 x \kappa_{cb}:\frac{j^b_{L,z}}{c_+} \frac{j^c_{L, \bar z}}{c_-}:(x) \cr
& = \frac{\epsilon}{4\pi f^4 c_+ c_-} \int d^2 x \left(
c_1 \frac{j^a_{L,\bar z}(x)}{(z-x)^2} + \tilde c j^a_{L,z}(x) 2\pi \delta^{(2)}(z-x) +... \right).
\end{align}
The ellipses contains higher-order terms both in $f^2$ and in the
distance between $z$ and $x$.  We will not keep track of these terms
for the time being, and we will comment on their relevance at the end
of the computation.  We now have to take the OPE of the previous
result with the primary field $\phi$. We obtain :
\begin{align}\label{jMargPhi} \frac{\epsilon}{4\pi f^4 c_+ c_-}& \int d^2 x \left( 
- \frac{c_1 c_-}{c_++c_-} t^a \frac{\phi(w)}{(z-x)^2(\bar x - \bar w)}
- \frac{\tilde{c} c_+}{c_++c_-} t^a \frac{\phi(w)}{x-w} \delta^{(2)}(z-x) + ... \right) \cr
& = \frac{\epsilon}{2 f^4 c_+ c_-}\left(\frac{c_1 c_-}{c_++c_-}-\frac{\tilde{c} c_+}{c_++c_-}\right) t^a \frac{\phi(w)}{z-w}+ ... \cr
& = -\epsilon c_+ t^a \frac{\phi(w)}{z-w}+ \text{less singular,}
\end{align}
where we used the explicit value of the coefficients \eqref{candg}. 
As claimed, the structure of the OPEs \eqref{defPrimariesBis} is
unaltered after perturbation of the kinetic term.  It is also
straightforward to check that (taking into account the renormalization
of the currents) the perturbation $f^2 \to f^2 + \epsilon$ induces a
deformation of the coefficients in \eqref{defPrimariesBis} that
matches the result obtained at first order in $\epsilon$.

Now let us come back to the terms we discarded in equation
\eqref{jMarg}. They contain the contribution to this computation from
the poles and less singular terms in the current algebra
\eqref{euclidOPEs}. All these terms are (composites of) currents. It
follows from \eqref{defPrimariesBis} and from dimensional analysis
that in the OPE between any one of these terms and the primary field
$\phi$, the most singular term that may arise multiplies the operator
$\phi$. Here we assume that all terms appearing in the OPE
\eqref{defPrimariesBis} can be written as composites of currents with
the field $\phi$. Thus if any of these terms has any effect on the
previous computation, it may at worse modify the coefficient obtained
in \eqref{jMargPhi}. On the other hand, as was mentioned in section
\ref{primaries}, the coefficients in \eqref{defPrimariesBis} are
fixed by demanding compatibility with current conservation and the
Maurer-Cartan equation. Since these coefficients were already
recovered in \eqref{jMargPhi} it follows that the term we discarded in
equation \eqref{jMarg} indeed has no effect on the result of the
computation. This can also be checked by hand for the terms that are
explicitly given in equation \eqref{euclidOPEs}.

\subsection{Current-primary OPE at order $f^2$}\label{AppjPhi}

Equation \eqref{defPrimaries} gives the OPE between a current and a
primary field at leading order.  According to the discussion of
section \ref{bootstrap} it is possible to compute 
the higher-order terms thanks to current conservation and the
Maurer-Cartan equation. In this appendix we perform the computation of
the first correction to the OPE \eqref{defPrimaries}, which leads to
the OPE \eqref{jPhiO1} in the bulk of the
 paper.

The terms on the right-hand side of the OPE \eqref{defPrimaries} are
of order $f^0$. 
We will now compute the current-primary OPE at
order $f^2$.  Following the discussion of appendix \ref{XXOPEs} we
make the following educated ansatz for the OPEs between the
left-currents and a primary field $\phi$:
\begin{align}
j^a_{L,z}(z) \phi(w) = & -\frac{c_+}{c_++c_-} \frac{t^a \phi(w)}{z-w} + :j^a_{L,z} \phi:(w) \cr
& + {A^a}_c \log|z-w|^2 :j^c_{L,z} \phi:(w) + {B^a}_c \frac{\bar z - \bar w}{z-w} :j^c_{L,\bar z} \phi:(w) + \mathcal{O}(f^4) \cr
j^a_{L,\bar z}(z) \phi(w) = & -\frac{c_-}{c_++c_-} \frac{t^a \phi(w)}{\bar z-\bar w} + :j^a_{L,\bar z} \phi:(w) \cr
& + {D^a}_c \log|z-w|^2 :j^c_{L,\bar z} \phi:(w) + {C^a}_c \frac{z-w}{\bar z - \bar w} :j^c_{L, z} \phi:(w) + \mathcal{O}(f^4).
\end{align}
We expect the coefficients ${A^a}_c$, ${C^a}_c$, ${B^a}_c$, ${D^a}_c$
to be of order $f^2$. We will check that the coefficient of the
first-order poles are not modified.  
As explained in section \ref{bootstrap} the demand of consistency with current conservation \eqref{phiCC} imposes that the terms in the $j^a_{L,\bar z}(z) \phi(w)$ OPE can be deduced from the terms in the $j^a_{L,z}(z) \phi(w)$:
\be {A^a}_c + {C^a}_c = 0 = {B^a}_c + {D^a}_c. \ee
To get further constraints on the tensors ${A^a}_c$ and ${B^a}_c$ we ask for the vanishing of the first-order poles in the OPE between the operator
$\phi$ and the Maurer-Cartan operator, that we write as in \eqref{phiModMC}:
\be\label{MC.phi=0} [ \bar \p j^a_{L,z}(z)  + i f^2 {f^a}_{bc} :j^c_{L,z} j^b_{L,\bar z}:(z)]\phi(w) = 0. \ee
The first part of this OPE is:
\begin{align} \bar \p j^a_{L,z}(z) \phi(w) = &  {A^a}_c  \frac{:j^c_{L,z}\phi:(w)}{\bar z - \bar w}  + {B^a}_c  \frac{:j^c_{L,\bar z}\phi:(w)}{z -w} +  \mathcal{O}(f^4).
\end{align}
The simple poles in the previous expression should be canceled by the
simple poles in the OPE between the composite operator $ i f^2 {f^a}_{bc}
:j^c_{L,z} j^b_{L,\bar z}:$ and the operator $\phi$. 
Notice that because of the factors $f^{2}$ multiplying the
composite operator, we only need to compute the OPE at order $f^0$.
We calculate :
\begin{align}
\phi(w)&[ i f^2 {f^a}_{bc} :j^c_{L,z} j^b_{L,\bar z}:(z)] =   i f^2 {f^a}_{bc} \lim_{:x \to z:} \phi(w)j^c_{L,z}(x) j^b_{L,\bar z}(z) \cr
= &   i f^2 {f^a}_{bc} \lim_{:x \to z:} \left \{
\left[ - \frac{c_+}{c_++c_-} \frac{t^c \phi(w)}{x-w} + :j^c_{L,z} \phi:(w) 
+
...  \right] j^b_{L,\bar z}(z) \right. \cr
& \quad \left. + j^c_{L,z}(x) \left[ -\frac{c_-}{c_++c_-}\frac{t^b \phi(w)}{\bar z - \bar w} + :j^b_{L,\bar z} \phi:(w) + 
...\right] \right\} 
\end{align}
To proceed according to the prescription of appendix
\ref{compositeOPEs} we have to expand the fields in the first line
(respectively the second line) in the neighborhood of the point $x$
(respectively $z$). Then we have to perform the remaining OPEs between
the currents and the (derivatives of) the primary field $\phi$. Notice
however that all the terms proportional to ${f^a}_{cb} t^b t^c =
\frac{i}{2} {f^a}_{cb} {f^{bc}}_d t^d$ do vanish. Only the
regular term in the current-primary OPE will contribute to the result
at order $f^0$. 
Moreover it is straightforward to check that the terms proportional to ${A^a}_c$ and ${B^a}_c$ in the previous OPE do not contribute at order $f^0$.
We obtain:
\be
\phi(w)[ i f^2 {f^a}_{bc} :j^c_{L,z} j^b_{L,\bar z}:(z)]
=  -i f^2 {f^a}_{bc} \left(  \frac{c_+}{c_++c_-} \frac{t^c :j^b_{L,\bar z}\phi:(z)}{z-w}+ \frac{c_-}{c_++c_-} \frac{t^b :j^c_{L, z}\phi:(z)}{\bar z-\bar w} + ... 
\right).
\ee
where the ellipses contains terms of order $f^4$ as well as terms of
order zero in the distance between $z$ and $w$.  Gathering terms, we
conclude that we have the equalities:
\begin{eqnarray} {A^a}_c &=& \frac{c_-}{(c_++c_-)^2} i {f^a}_{cb} t^b + \mathcal{O}(f^4)
\nonumber \\
 {B^a}_c &=& \frac{c_+}{(c_++c_-)^2} i {f^a}_{cb} t^b + \mathcal{O}(f^4).
\end{eqnarray}
We note that one can reach the same conclusion by computing the OPE
between a current and both sides of the equation \eqref{dPhi=JPhi},
i.e. by demanding compatibility with the proportionality relation
between the operators $\p \phi$ and $ t_a :j^a_{L,z} \phi:$.

\subsection{Stress-tensor-primary OPE at order $f^2$}\label{AppTphi}

Here we present the computation of the OPE between the stress-energy
tensor and a primary field $\phi$. This computation relies on the
prescription of appendix \ref{compositeOPEs}, and on the 
current-current and
current-primary OPEs \eqref{euclidOPEs} and \eqref{jPhiO1}.
Since we computed these OPEs up
to order $f^2$, we will also obtain the stress-tensor OPE up to order
$f^2$.
\begin{align}\label{phiT}
\phi(z) 2 c_1 T(w)
&= \lim_{:x \to w:}\phi(z) j^a_{L,z}(x) j^b_{L,z}(w) \kappa_{ba} \cr
&= \kappa_{ab} \lim_{:x \to w:} \left[ \left(
-\frac{c_+}{c_++c_-} \frac{t^a \phi(z)}{x-z} + :j^a_{L,z} \phi:(x) \right. \right. \cr
& \left. + {A^a}_c \log |z-x|^2 :j^c_{L,z}\phi:(x) + {B^a}_c\frac{\bar z - \bar x}{z-x}:j^c_{L,\bar z}\phi:(x) + ... 
\right) j^b_{L,z}(w) \cr
& + \left. j^a_{L,z}(x) \left( -  \frac{c_+}{c_++c_-} \frac{t^b \phi(w)}{w-z} + \mathcal{O}\left( (z-w)^0 \right)
\right) \right]
\end{align}
Let us first consider the first term in the previous
expression. According to the prescription given in appendix
\ref{compositeOPEs}, we have to evaluate the operator $\phi(z)$ at the
point $x$ before we take the OPE with the remaining current
$j^b_{L,z}(w)$. So we rewrite this term as:
\begin{align} \kappa_{ab} & \lim_{:x \to w:} \left( -\frac{c_+}{c_++c_-} \frac{t^a}{x-z} \sum_{n,\bar n=0}^{\infty} \frac{(z-x)^n}{n!}\frac{(\bar z-\bar x)^{\bar n}}{\bar n!} \p^n \bar \p^{\bar n} \phi(x) \right) j^b_{L,z}(w) \cr
& =  \kappa_{ab} t^a \frac{c_+}{c_++c_-} \lim_{:x \to w:}\sum_{n,\bar n=0}^{\infty}\frac{(z-x)^{n-1}}{n!}\frac{(\bar z-\bar x)^{\bar n}}{\bar n!} \p_x^n \bar \p_x^{\bar n} \left(
-\frac{c_+}{c_++c_-} \frac{t^b \phi(x)}{w-x} - :j^b_{L,z} \phi:(w) + ...
\right) \cr
& =  -\kappa_{ab} t^a t^b \left(\frac{c_+}{c_++c_-} \right)^2
\lim_{:x \to w:}\sum_{n,\bar n=0}^{\infty}\frac{(z-x)^{n-1}}{n!}\frac{(\bar z-\bar x)^{\bar n}}{\bar n!} \p_x^n \bar \p_x^{\bar n}  \cr
& \qquad
\left(\frac{1}{w-x} \sum_{m,\bar m=0}^{\infty}\frac{(x-w)^{m}}{m!}\frac{(\bar x-\bar w)^{\bar m}}{\bar m!} \p^m \bar \p^{\bar m}\phi(w) 
\right) -   \kappa_{ab} t^a \frac{c_+}{c_++c_-} \frac{:j^b_{L,z} \phi:(w)}{w-z} + ... \nonumber
\end{align}
In the previous lines we only kept track of the operators that will lead to poles in the final result.
We evaluated the operator $\phi$ at the point $w$ so that the action of the derivatives is easier to take care of:
\begin{align}
 = -\kappa_{ab}& t^a t^b \left(\frac{c_+}{c_++c_-} \right)^2
\lim_{:x \to w:}\sum_{n,\bar n=0}^{\infty}\frac{(z-x)^{n-1}}{n!}\frac{(\bar z-\bar x)^{\bar n}}{\bar n!}
  \cr
& 
\left( \sum_{m,\bar m=0}^{\infty} (-1) \frac{(x-w)^{m-n-1}}{m\ (m-n-1)!} \frac{(\bar x - \bar w)^{\bar m - \bar n}}{(\bar m - \bar n)!}\p^m \bar \p^{\bar m}\phi(w) 
\right)-   \kappa_{ab} t^a \frac{c_+}{c_++c_-} \frac{:j^b_{L,z} \phi:(w)}{w-z} + ... \nonumber
\end{align}
The regular limit gives a non-zero result for the anti-holomorphic
factor only if $\bar m - \bar n = 0$. For the holomorphic factor, one
needs $m-n-1=0$. Notice that the terms with $n=m=0$ also contributes
with a non-vanishing term. Eventually we obtain:
\begin{align}
 \kappa_{ab}& t^a t^b \left(\frac{c_+}{c_++c_-} \right)^2
\left(  \sum_{n,\bar n=0}^{\infty}\frac{(z-w)^{n-1}}{(n+1)!}\frac{(\bar z-\bar x)^{\bar n}}{\bar n!} \p^{n+1} \bar \p^{\bar n}\phi(w) \right. \cr
& \qquad
\left.  + \sum_{\bar n=0}^{\infty}\frac{1}{(z-w)^2} \frac{(\bar z-\bar x)^{\bar n}}{\bar n!} \bar \p^{\bar n}\phi(w) 
  \right)-   \kappa_{ab} t^a \frac{c_+}{c_++c_-} \frac{:j^b_{L,z} \phi:(w)}{w-z} + ...\cr
 & =  \kappa_{ab} t^a t^b \left(\frac{c_+}{c_++c_-} \right)^2 \frac{\phi(z)}{(z-w)^2} -   \kappa_{ab} t^a \frac{c_+}{c_++c_-} \frac{:j^b_{L,z} \phi:(w)}{w-z} + ...
 \end{align}
 This completes the evaluation of the first term in \eqref{phiT}.  The
 other terms are much easier to deal with.  The only non-trivial part
 is the computation the OPE between a current $j^b_{L,z}(w)$ and the
 composite operators $:j^c_{L,z}\phi:(z)$ and $:j^c_{L,\bar
   z}\phi:(z)$. Since the coefficients ${A^a}_c$ and ${B^a}_c$ already
 are of order $f^2$, we only need to know these OPEs at order $f^0$.
 We find:
\begin{align} j^b_{L,z}(w):j^c_{L,z}\phi:(z) = &
\left(c_1 \kappa^{bc} + \frac{c_-(c_2-g)-c_+c_2}{c_++c_-}{f^{bc}}_d t^d \right)\frac{\phi(z)}{(w-z)^2} \cr
& \quad - \frac{c_+}{c_++c_-} \frac{t^b :j^c_{L,z}\phi:(z)}{w-z} + \mathcal{O}(f^2) \\
j^b_{L,z}(w):j^c_{L,\bar z}\phi:(z) = &
\left(\tilde{c} \kappa^{bc} + \frac{c_-(c_2-g)+c_+(c_4-g)}{c_++c_-}{f^{bc}}_d t^d \right)\phi(z)2\pi \delta^{(2)}(w-z) \cr
& \quad - \frac{c_+}{c_++c_-} \frac{t^b :j^c_{L,\bar z}\phi:(z)}{w-z} + \mathcal{O}(f^2). \end{align}
All the terms that appear in these OPEs give zero once contracted
either with ${A^a}_c \kappa_{ab}$ or with ${B^a}_c \kappa_{ab}$.  In
particular factors of the form ${f^a}_{bc}t^c t^b$ vanish since the
dual Coxeter number is zero.
Gathering everything we obtain:
\begin{align}
\phi(z) 2 c_1 T(w)
&= \frac{c_+^2}{(c_++c_-)^2}t^a t^b \kappa_{ab} \frac{\phi(z)}{(z-w)^2} -\frac{2 c_+}{c_++c_-} \frac{\kappa_{ab}t^a :j^b_{L,z}\phi:(z)}{w-z}+ \mathcal{O}(z-w)^{0}+ \mathcal{O}(f^2).
\end{align}
The previous result is true only up to terms of order $f^2$, since a
term of order $f^4$ in the current-primary OPE may give a term of
order $f^2$ once contracted with an additional current (see lemma \eqref{lemmaf-2bis}). We rewrite the
result as:
\begin{align}
T(w) \phi(z)
&= \frac{f^2}{2} \frac{t^a t^b \kappa_{ab} \phi(z)}{(z-w)^2} +\frac{1}{c_+} \frac{\kappa_{ab}t^a :j^b_{L,z}\phi:(z)}{w-z}+ \mathcal{O}(z-w)^{0}+ \mathcal{O}(f^4) 
\end{align}
This concludes the proof of equation \eqref{Tphi} in section \ref{primaries}.

\section{The mode expansion on the cylinder}
\label{commutators}
When the theory is defined on a cylinder we can expand the operators
in modes by means of a Fourier transform along the compact coordinate.
Then we can convert the current-current OPEs into graded commutation
relations for the modes of the currents.  This was done for the
current algebra \eqref{euclidOPEs} in \cite{Ashok:2009xx}.  In this
appendix we give the translation of the left current - right current
OPEs (\ref{jLjR1}, \ref{jLjR2}) in terms of commutation relations.  We
use the same techniques as in section 5 of \cite{Ashok:2009xx}. To
simplify the notation we do not write explicitly the subscript $L$ or $R$
on the currents since it is redundant with the different notation for
the left and right adjoint representations. We expand the currents and
the adjoint operator in modes:
 \beq j^a_{z}(\sigma,\tau) &=& +i
\sum_{n \in Z} e^{-in\sigma}j^a_{z,n}(\tau) \cr j^a_{\bar
  z}(\sigma,\tau) &=& -i \sum_{n \in Z} e^{-in\sigma}j^a_{\bar
  z,n}(\tau) \cr j^{\bar a}_{z}(\sigma,\tau) &=& +i \sum_{n \in Z}
e^{-in\sigma}j^{\bar a}_{z,n}(\tau) \cr j^{\bar a}_{\bar
  z}(\sigma,\tau) &=& -i \sum_{n \in Z} e^{-in\sigma}j^{\bar a}_{\bar
  z,n}(\tau) \cr
 \mathcal{A}^{a \bar a}(\sigma,\tau) &=& \sum_{n \in Z}
e^{-in\sigma} \mathcal{A}^{a \bar a}(\tau).
\eeq We obtain the
commutation relations: \beq [j^a_{z,n},j^{\bar a}_{z,m}] &=&
+\frac{c_+c_-}{c_++c_-} \frac{c_-m-c_+ n}{c_++c_-} \mathcal{A}^{a \bar
  a}_{n+m} \cr [j^a_{\bar z,n},j^{\bar a}_{\bar z,m}] &=& -
\frac{c_+c_-}{c_++c_-} \frac{c_+m-c_-n}{2} \mathcal{A}^{a \bar a}_{n+m} \cr
[j^a_{z,n},j^{\bar a}_{\bar z,m}] &=& -\frac{c_+^2 c_-}{(c_++c_-)^2}
(m+n) \mathcal{A}^{a \bar a}_{n+m} \cr [j^a_{\bar z,n},j^{\bar a}_{z,m}] &=&
+\frac{c_+c_-^2}{(c_++c_-)^2} (m+n) \mathcal{A}^{a \bar a}_{n+m}, \eeq as
well as the standard commutation relations between the modes of the
currents and the left-right adjoint primary (as determined by their
OPE), and the left-left commutation relations calculated in
\cite{Ashok:2009xx}.

In \cite{Ashok:2009xx} it was shown that the combination of left
current components $j^a_{z,n}-j^a_{\bar z,n}$ generate a Kac-Moody
algebra at integer level $k$. This is also the case for the right
combination $j^{\bar a}_{z,m}-j^{\bar a}_{\bar z,m}$.  As a
consequence of the above commutation relations, we find moreover that
the left and right Kac-Moody subalgebras commute: \be
[j^a_{z,n}-j^a_{\bar z,n},j^{\bar a}_{z,m}-j^{\bar a}_{\bar z,m}] = 0.
\ee Only the zero modes of these affine currents commute with the
worldsheet Hamiltonian.

\section{Classical integrability}
\label{classint}
In this appendix, we will show that 
principal chiral models with or without Wess-Zumino term are
classically integrable. We generalize here the standard calculation to the case
with non-zero Wess-Zumino term.
The equations of motion $d \ast j = 0$ for the model written in terms of the
left current components read:
\begin{eqnarray}
\bar{\partial} j^a_z + \partial j^a_{\bar{z}} &=& 0,
\end{eqnarray}
where we have that:
\begin{eqnarray}
j_z &=& - \frac{1}{2} ( \frac{1}{f^2} + k) \partial g g^{-1}
\nonumber \\
j_{\bar{z}} &=& - \frac{1}{2} ( \frac{1}{f^2} - k) \bar{\partial} g g^{-1}.
\end{eqnarray}
As before, the coefficient of the principal chiral model term is
$1/f^2$ and the Wess-Zumino term has coefficient $k$.  The
Maurer-Cartan equation $d (dg g^{-1} ) = dg g^{-1} \wedge dg g^{-1}$
is:
\begin{eqnarray}
 - \frac{1}{2} ( \frac{1}{f^2} - k) \bar{\partial} j^a_z +  \frac{1}{2} ( \frac{1}{f^2} + k) \partial j^a_{\bar{z}} - i {f^{a}}_{bc} j_z^c j_{\bar{z}}^b &=& 0. 
\end{eqnarray}
In this context it is easier to work with the canonical right invariant one-form:
\begin{eqnarray}
\omega &=& dg g^{-1}
\end{eqnarray}
and rewrite the equations of motion in terms of $\omega$ and the coefficients
$c_\pm$ defined as in the bulk of the paper:
\begin{eqnarray}
 \bar{\partial} \omega_z &=& -\frac{ c_-}{c_+ + c_-} [\omega_z , \omega_{\bar{z}}]
\nonumber \\
\partial \omega_{\bar{z}} &=& + \frac{c_+}{c_++c_-} [\omega_z , \omega_{\bar{z}}].
\label{EOMs}
\end{eqnarray}
Now consider a connection which is a function of a spectral parameter $\lambda$:
\begin{eqnarray}
A(\lambda) &=& -\frac{2}{1+\lambda} \frac{c_+}{c_++c_-} \omega_z dz - 
\frac{2}{1-\lambda} \frac{ c_-}{c_++c_-} \omega_{\bar{z}} d \bar{z}
\end{eqnarray}
and compute the curvature of the connection:
\begin{eqnarray}
F_{\bar{z} z} &=& -\frac{2}{1+\lambda}  \frac{c_+}{c_++c_-}  \bar{\partial} \omega_z +
 \frac{2}{1-\lambda}  \frac{ c_-}{c_++c_-}  \partial \omega_{\bar{z}} -
 \frac{c_+}{c_++c_-}  \frac{ c_-}{c_++c_-}  2 (\frac{1}{1+\lambda}+  \frac{1}{1-\lambda}) 
[\omega_z ,\omega_{\bar{z}}]. \nonumber
\end{eqnarray}
Flatness of the connection for all values of the spectral parameter $\lambda$
is equivalent to the validity of the equations of motion (\ref{EOMs}).
Using the on-shell flat connection, we can
define an infinite set of conserved charges, for instance by calculating the 
traced holonomy
for the model on a circle times time, and expanding in the spectral parameter.
The infinite set of conserved charges renders the theory classically integrable. The theory can 
then be studied using the powerful tools of integrability.

\end{appendix}

\end{document}